\documentclass{aa}  

\usepackage{graphicx}
\usepackage{txfonts}
\usepackage[toc,page]{appendix}
\usepackage{color}

\begin{document}

\title{Galaxy clusters in the context of superfluid dark matter}

   \subtitle{}

   \author{Alistair O. Hodson$^1$
          \and
          Hongsheng Zhao$^1$
          \and
          Justin Khoury$^2$
          \and
          Benoit Famaey$^3$
          }

   \institute{$^1$ School of Physics and Astronomy, University of St Andrews,
             Scotland \\
            $^2$ Center for Particle Cosmology, Department of Physics and Astronomy, University of
            Pennsylvania, Philadelphia, PA 19104, USA \\
             $^3$ Universit\'e de Strasbourg, CNRS UMR 7550, Observatoire astronomique de Strasbourg, 11 rue de l'Universit\'e, 67000 Strasbourg, France \\
              \email{aoh2@st-andrews.ac.uk}
              }

%   \date{Received September 15, 1996; accepted March 16, 1997}

% \abstract{}{}{}{}{} 
% 5 {} token are mandatory
 
  \abstract
   {The mass discrepancy
in the Universe has not been solved by the cold dark matter (CDM)
or the modified Newtonian dynamics (MOND) paradigms so far. The
problems and solutions of either scenario are mutually exclusive
on large and small scales. It has recently been proposed, by assuming that dark matter is a superfluid, that MOND-like effects can be achieved on small scales whilst preserving the success of $\Lambda$CDM on large scales. Detailed models within this ``superfluid dark matter" (SfDM) paradigm are yet to be constructed.}
   {Here, we aim to provide the first set of spherical models of galaxy clusters in the context of SfDM. We aim to determine whether the superfluid formulation is indeed sufficient to explain the mass discrepancy in galaxy clusters.}
   {\textcolor{black}{The SfDM model is defined by two parameters. $\Lambda$ can be thought of as a mass scale in the Lagrangian of the scalar field that effectively describes the phonons, and it acts as a coupling constant between the phonons and baryons.  $m$ is the mass of the DM particles. Based on these parameters, we outline the theoretical structure of the supefluid core} and the surrounding ``normal-phase" dark halo of quasi-particles. The latter are thought to encompass the largest part of galaxy clusters. Here, we set the SfDM transition at the radius where the density and pressure of the superfluid and normal phase coincide, neglecting the effect of phonons in the superfluid core. We then apply the formalism to a sample of galaxy clusters, and directly compare the SfDM predicted mass profiles to data.}
   {We find that the superfluid formulation can reproduce the X-ray dynamical mass profile of clusters reasonably well, but with a slight under-prediction of the gravity in the central regions. This might be partly related to our neglecting of the effect of phonons in these regions. Two normal-phase halo profiles are tested, and it is found that clusters are better defined by a normal-phase halo resembling an Navarro-Frenk-White-like structure than an isothermal profile.}
   {In this first exploratory work on the topic, we conclude that depending on the amount of baryons present in the central galaxy and on the actual effect of phonons in the inner regions, this superfluid formulation could be successful in describing galaxy clusters. In the future, our model could be made more realistic by exploring non-sphericity and a more realistic SfDM to normal phase transition. The main result of this study is an estimate of the order of magnitude of the theory parameters for the superfluid formalism to be reasonably consistent with clusters. These values will have to be compared to the true values needed in galaxies.}


   \maketitle

\section{Introduction}

While observations of large-scale structure in the Universe are traditionally explained by invoking a non-uniform distribution of dark matter (DM) particles,  the nature of such particles and the strength of their interactions so far remain unknown. Such particles might be of any type as long as they are collision-free non-relativistic massive particles on scales of galaxy clusters and above, as in the standard $\Lambda$CDM paradigm. A strong argument for the existence of such particles is the colliding bullet cluster \citep{bullet}, which has a lensing signature offset from its baryonic centre. This is explained in the \textcolor{black}{cold dark matter (CDM)} paradigm, as the gas of the cluster interacts while CDM does not, and thus the gas stays closer to the centre and the dark matter can pass through almost unaffected. This therefore produces the observed lensing signature of the bullet cluster. 

{ Until DM particles are directly detected, it is useful, however, to keep in mind that its properties could be more complex than currently envisaged \citep[e.g.][]{BEDM,fuzzyDM}.} For instance, on small scales, there are still some issues with interaction-free particle models, especially to explain the  dynamics of galaxies. The most widely discussed problems are the cusp-core problem, the
missing satellites problem, the too-big-to-fail problem, and satellite plane problems \citep[e.g.][]{cuspcore1, cuspcore2,deblokCoreCusp, missing1, missing2, TBTF1, TBTF2, plane1,Pawlowski}. These open
questions have led a part of the community to search for an alternative
to the \textcolor{black}{CDM} paradigm to explain the observed dynamics of galaxies. The general problem of CDM in galaxies indeed seems to be more profound than the series of problems listed above. In particular, a diversity of shapes of rotation curves at a given maximum circular velocity scale are observed, which is contrary to CDM expectations \citep{Oman}, and a uniformity of shapes exists at a given baryonic surface density scale \citep{Famaey}. These puzzling observations can be summarised by the radial acceleration relation \citep{McGaugh}, which is equivalent to the well-known
modified Newtonian dynamics (or MOND) phenomenology for isolated rotationally supported systems \citep{milgrom19832,milgrom19833, milgrom19831, bekenstein1984, Famaey}. The idea of MOND was to introduce an acceleration scale $a_{0}$ in the dynamics, directly illustrating the role that is played by the baryonic surface density in observations. When the gravitational acceleration is much higher than $a_{0}$, the gravity behaves as Newton predicts. Much below $a_{0}$, the force law of gravity effectively switches to a force law proportional to $1/r$ (instead of Newton's $1/r^{2}$ \textcolor{black}{law}). This is also the actual phenomenology of the radial acceleration relation, or mass discrepancy acceleration relation (MDAR). The recent study of \cite{MDAR1} aims to determine whether the MDAR can be naturally predicted by $\Lambda$CDM. It was found that with a mass-dependent DM profile, the MDAR
can be reproduced, but a universal Navarro-Frenk-White (NFW) profile does not work well in systems below $M_{\star} \approx 10^{9.5} M_{\odot}$. { Other studies of this problem found that $\Lambda$CDM  was able to predict the  general trend of the MDAR \citep{Keller2017,Navarro2016,Ludlow2017}, but whether the observed scatter and normalisation can be precisely reproduced is still fiercely debated \citep{MDAR2,LelliBTFR}. On the other hand, a very small scatter would be expected in MOND by construction. However, the MOND phenomenology fails on large scales}, for example, galaxy clusters \citep[see for example][]{sanders1999}, although work has been conducted to reconcile this problem by including neutrinos \citep[e.g.][]{sanders2003,angus20081,angus2009} or by modifying the MOND formulation itself \citep{EMOND,HodsonEMOND,HodsonUDG}. 

In short, $\Lambda$CDM is successful on large scales, but still has some currently debated problems on small (galactic and subgalactic) scales, whilst MOND seems to have the exact opposite problem. It is therefore interesting to ask the question whether we can have MOND behaviour on small scales and CDM-like behaviour on larger scales. Following up on previous proposals by \citet{Blanchet},
for instance, this is exactly what Khoury and Berezhiani recently explored in a series of papers~\citep{SF2,SF3,SF4}. The idea expressed in these works is the concept that DM behaves like a superfluid in cold enough and dense enough environments, typically within galaxies, and it behaves like normal particle DM (``normal phase'') in clusters and on larger scales. This framework aims to describe the rotation curve in galaxies through a MOND-like phonon-mediated force resulting from DM when it is in its superfluid phase. In galaxy clusters, however, most of the matter is outside the superfluid phase and is made of quasi-particles in thermal equilibrium.

Each paper discussed a different approach to achieve the desired effect, and we focus here on~\citet{SF2} and~\citet{SF3}. These two papers prescribe the fundamental components to the theory, but attempts to make a detailed model of an astrophysical system
are lacking so far. Further to the work in these two papers, we must also address how to model the normal phase of the matter and how that embeds the superfluid. In a first attempt to do this, we consider the case of spherical galaxy clusters. The reasons for first choosing galaxy clusters instead of galaxies are that galaxy clusters are commonly modelled in spherical symmetry without the need for a disk component, and very little contribution from the phonon force is therefore expected. 

Section~\ref{model} outlines how we modelled the dark matter, and we describe the superfluid phase, the normal phase, the transition
of one phase transitions to the other, and the determination of the virial mass and radius of the cluster in this context. In Sect.~\ref{toy} we describe a toy model with no baryons for illustrative purposes. In Sect.~\ref{dynamical} we then analyse a sample of galaxy clusters in the context of this model, mainly comparing the derived mass profile of the superfluid to the profile calculated from hydrostatic equilibrium of the gas. We conclude in Sect.~\ref{conclusion}.

\section{Dark matter superfluid model}\label{model}

In this section we give a brief overview of the equations that describe a system in the context of the DM superfluid theory \textcolor{black}{(SfDM)}. We first recall the general idea of the SfDM model. Ignoring interactions for simplicity, DM particles undergo a phase transition to the superfluid state whenever their de Broglie wavelength $\lambda_{\rm dB} \sim 1/mv$ is larger than the average interparticle separation $\ell\sim n^{-1/3}$. Here $v$ is typically the DM velocity dispersion, $m$ is the DM particle mass, and $n$ is the local average DM number density. Thus superfluidity arises in sufficiently cold (large $\lambda_{\rm dB}$) and dense (small $\ell$) environments. Another requirement for Bose-Einstein condensation is that DM reaches thermal equilibrium, which requires sufficiently strong self-interactions~\citep{SF1}.

We expect the superfluid phase to be confined within the central regions of clusters, where the density is high enough. Because clusters are hotter than galaxies, most of their DM content should be in the normal phase. The DM in galaxy clusters therefore consists of a superfluid core, whose radius (depending on the theory parameter values) will range from \textcolor{black}{ approximately 50 to 100~kpc}, surrounded by an atmosphere of DM particles in the normal phase. The key element to model here is the transition from one phase to the other. If too large a region in the cluster is in the superfluid phase, it is likely that observed X-ray gas temperature profiles cannot be reproduced.

Within the superfluid core, DM is more aptly described as collective excitations, which at low energy are phonons. In the superfluid paradigm of~\citet{SF2,SF3}, the MOND-like effects are achieved as a result of phonon excitations in the superfluid phase mediating a long-range force between ordinary matter particles. This phonon-mediated force is only important at low acceleration ($a\ll a_{0}$), resulting in strong deviations from Newtonian gravity. In galaxies, this effect is critical in reproducing the empirical success of MOND at fitting rotation curves. The central regions of galaxy clusters, however, tend to lie in the intermediate ($a \sim a_{0}$) to Newtonian regime ($a\gg a_{0}$), where the phonon-mediated force is at most comparable to the Newtonian force. \textcolor{black}{We therefore expect that the phonon force is unimportant in galaxy clusters compared to the DM component and therefore, in the interest of simplicity, we assume that the phonon contribution to our calculations is hereafter zero}. \textcolor{black}{We stress that this is in no way an acceptable assumption when the internal accelerations are small compared to $a_{0}$ , and thus some of the equations described in the following sections cannot be applied to galaxies. The phonon force must be included for consistency.} 

\subsection{Set Up}

The gravitational potential $\Phi$ is determined as usual by Poisson's equation,
\begin{equation}
\frac{1}{4 \pi G_{\rm N}}\nabla^2 \Phi(\vec{r}) =\rho(\vec{r}) + \rho_{\rm b}(\vec{r}) \,,
\end{equation}
where $\rho_{\rm b}$ and $\rho$ denote the baryon and DM densities, respectively. This allows us to integrate for $\Phi$ for any given equation of state $\rho(\Phi)$.  

For simplicity, we assume hydrostatic equilibrium, which requires 
\begin{equation}
\label{pressuredensity}
P(\vec{r}) =  \int^{\infty}_{\vec{r}} \rho \vec{\nabla} \Phi \cdot \vec{{\rm d}r}\,,
\end{equation}
where $\rho$ and $P$ are the mass density and pressure of the DM, respectively.
This allows us to obtain the pressure everywhere.  

In the following, we furthermore assume spherical symmetry,  in which case the above equations reduce to 
\begin{equation}
\rho(r) + \rho_{\rm b}(r)    = \frac{1}{4 \pi G_{\rm N}r} \frac{{\rm d}^2 (r \Phi)}{{\rm d}r^2}\,;
\label{sphericalPoisson}
\end{equation}
and
\begin{equation}
\frac{{\rm d}P}{{\rm d}r} = - \rho \frac{{\rm d}\Phi}{{\rm d}r}\,.
\label{sphericalhydrostatic}
\end{equation}

To solve these equations and derive the DM profile in galaxy clusters, multiple parts of the model need to be addressed:
\begin{itemize}
\item DM superfluid core profile
\item DM normal halo profile
\item Matching the superfluid core to the normal halo
\item Determining the virial radius of the system
\item Making total mass the free parameter.
\end{itemize}  
These ingredients are described in turn below.

\subsection{Dark matter superfluid core}\label{DMsuperfluidphase}

We begin by reviewing some properties of the inner superfluid core, discussing the pressure, sound speed, gravitational potential, and how they are related.

Ignoring phonons, the pressure of the superfluid is related to its density $\rho_{\rm s}$ via~\citep{SF2}%
\begin{equation}\label{pressure2}
P_{\rm s} = \frac{\rho_{\rm s}^{3}}{12 K^{2}}\,; \hspace{5mm} K \equiv \frac{\Lambda c^{2} m^{3}}{\hbar^{3}} \,,
\end{equation}
where $\Lambda$ is a mass scale and $c$ is the speed of light. We chose $\Lambda$ to be a mass scale to simplify the units of the $\Lambda m^{3}$ combination that appears throughout. We present $\Lambda m^{3}$ in units of $eV^{4}/c^{8}$. However, we have made our equations dimensionally sound in real units such that $\Lambda$ has units of mass, m has units of mass, and $\hbar$ has normal units kg m$^{2}$ s$^{-1}$. The superfluid parameters $\Lambda$ and $m$ always appear in the combination $\Lambda m^3$ in our calculations, hence there exists a degeneracy when choosing values. We therefore combine the parameters in this way and only discuss the value of $K = \Lambda c^{2} m^{3}/\hbar^{3}$ in what follows. \textcolor{black}{The degeneracy of the parameter combination $\Lambda m^{3}$ is broken when the phonon force is included. Therefore, as phonons become important on small scales, applying the superfluid paradigm to galaxies will be key in fully understanding these parameters (Berezhiani et al. in prep.). }

Equation~\eqref{pressure2} describes a polytropic equation of state, $P \sim \rho^{1 + 1/n}$, with index $n = 1/2$. Substituting the superfluid equation of state in the equation of the hydrostatic equilibrium~Eq.~\ref{sphericalhydrostatic}, we obtain
\begin{equation}
\frac{\rho_{\rm s}}{4K^2} \frac{{\rm d}\rho_{\rm s}}{{\rm d}r} = -\frac{{\rm d}\Phi}{{\rm d}r}\,.
\end{equation}
It is easy to see that this is solved by
\begin{equation}
\rho_{\rm s}(r) = 2K \sqrt{-2 \Phi(r)}\,.
\label{superfluiddensity}
\end{equation}
Thus the superfluid density is uniquely specified once we know the gravitational potential. The latter is fixed by integrating Poisson's equation~(Eq.~\ref{sphericalPoisson}):
\begin{equation}\label{Poissonbary2}
\frac{1}{4 \pi G_{\rm N} r^{2}}\left( r^{2} \Phi'(r) \right)'= 2 K \sqrt{-2 \Phi(r)} + \rho_{\rm b}(r)\,.
\end{equation}
This can be solved with initial conditions, $\Phi(r=0) = \Phi_{0}$ and $\Phi'(r=0) = 0,$ where $\Phi_{0}$ is a free parameter to be determined. The central potential gradient should be set in accordance with the baryon potential gradient. However, setting our value to zero did not affect our results. We discuss how we determined this central potential in Sect.~\ref{massparam}. 

The adiabatic sound speed is as usual given by
\begin{equation}\label{soundspeed}
c_{s}^{2} = \frac{{\rm d}P_{\rm s}}{{\rm d}\rho_{\rm s}} = \left(\frac{\rho_{\rm s}}{2 K}\right)^{2} = -2\Phi(r)\,.
\end{equation}
Thus the superfluid sound speed $c_{s}$ is also related to the gravitational potential $\Phi$. 

In the absence of baryons ($\rho_{\rm b} = 0$), Poisson's equation~Eq.~\ref{Poissonbary2} can be cast as a Lane-Emden equation~\citep{SF2}. The resulting density profile is smooth at the origin and vanishes at a certain radius that defines the core radius. This radius is determined by the parameters of the model ($m$ and $\Lambda$) together with the central density. In the presence of baryons, the superfluid profile will of course be altered, but the characteristic feature of a superfluid profile vanishing at a particular radius will remain.

%Finally, we introduce the chemical potential of the superfluid. If we have a scenario in which there are two systems, each will have its own chemical potential. Particles will move between the systems such that particles with higher chemical potential will move to the lower chemical potential system. When the two systems have the same chemical potential, they are known to be in diffusive equilibrium. This is analogous to temperature balancing and thermal equilibrium. For our superfluid, the pressure can be written in terms of the chemical potential per unit mass, $\hat{\mu}/m$, as
%
%\begin{equation}\label{pressure1}
%P_{\rm s} = \frac{2 K}{3 }\left( \frac{2  \hat{\mu}}{m}  \right)^{3/2}\,.
%\end{equation}
%
%\noindent From Eq.~\ref{pressure2}, we see that the chemical potential can be written in terms of the superfluid density as
%
%\begin{equation}
%\frac{\hat{\mu}}{m} = \frac{\rho_{\rm s}^{2}}{8 K^{2}}= \frac{c_{s}^{2}}{2} = -\Phi(r)\,.
%\end{equation}
%
%\noindent Hence the chemical potential, sound speed and gravitational potential carry the same information. 

\subsection{Dark matter normal halo}

\subsubsection{Isothermal}

The superfluid core is assumed to be surrounded by an atmosphere of normal-phase DM particles in thermal equilibrium. For simplicity, we ignore interactions and treat this normal component as an ideal gas,
\begin{equation}
P_{\rm n} =\frac{k_{\rm B}T}{m} \rho_{\rm n} \,.
\end{equation}
Hydrostatic equilibrium in this case implies the well-known isothermal profile
\begin{equation}\label{rhonormal}
\rho_{\rm n} = \rho_0 \exp{\left( - \frac{m \Phi(r)}{k_{\rm B}T}\right)} \,.
\end{equation}
\noindent \textcolor{black}{In the simplified case where we omit the contribution of the baryons, substituting Eq. ~\ref{rhonormal} into the Poisson equation yields the simplified solution}
\begin{equation}\label{normalphasedensity}
\rho_{\rm n} =  \rho_c \left(\frac{R_c}{r}\right)^2\,.
\end{equation}
The density normalisation $\rho_c$ and radius $R_c$ are fixed below. 

Isothermal DM haloes have an enclosed mass that grows linearly with radius, resulting from the density $\rho \propto r^{-2}$.

\subsubsection{NFW halo}

The isothermal case described above is very simple in formulation. For galaxy clusters in $\Lambda$CDM, it is common to model haloes using an NFW profile \citep{NFWpaper,zhaoprofile} which has, at large radius, $\rho \propto r^{-3}$ and a logarithmic mass growth. To mimic this type of behaviour for our normal phase, we tested a density profile of the form

\begin{equation}\label{NFWDen}
\rho_{\rm n} = \rho_{\rm c} \frac{R_{\rm c}}{r}\left( 1+\frac{r}{r_{\rm s}} \right)^{-2}\left( 1+\frac{R_{\rm c}}{r_{\rm s}} \right)^{2}.
\end{equation}

Equation \ref{NFWDen} has the properties $\rho_{\rm n} = \rho_{\rm c}$ when $r=R_{\rm c}$, $\rho_{\rm n}\propto r^{-1}$ when $r\ll r_{\rm s}$ , and $\rho_{\rm n}\propto r^{-3}$ when $r\gg r_{\rm s}$. { Here the scale radius $r_s$ was \textcolor{black}{estimated} to closely match NFW profiles for each cluster.} For the first toy model examples in Sect.~3 below, we demonstrate the equations with the isothermal profile only for simplicity, but we show the NFW result when analysing the clusters in Sect.~4. 

\subsection{DM in two phases}

Ideally, we would like to build a model with three sections: 1)~the inner section, which is dominated by matter in the superfluid phase; 2)~the outer halo, which is dominated by normal phase matter; and 3)~a transition regime, which has a mixture of normal-phase and superfluid-phase particles. We neglect this third regime in our model for simplicity and design a system where below the core radius $R_{\rm c}$  only superfluid matter exists, and above $R_{\rm c}$ only normal-phase matter is present. The procedure for determining this radius involves continuity of pressure between the two phases of DM and is outlined in Sect.~\ref{corecalc}. 

In other words, we modelled the DM in galaxy clusters as a superfluid core ($r \leq R_c$) with density $\rho_{\rm s}  = 2K \sqrt{-2 \Phi(\vec{r})}$ and pressure $P_{\rm s} = \frac{\rho^3_{\rm s}}{12K^2}$, { surrounded by a normal phase of DM in the outer halo ($r \geq R_c$) with a density profile following Eq.~\ref{normalphasedensity} for the isothermal case or Eq.~\ref{NFWDen} for the NFW case} and pressure $P_{\rm n} =\frac{k_{\rm B}T}{m} \rho_{\rm n}$. The gravitational potential $\Phi(r)$ has a meaningful zero-point fixed by the central density of the superfluid, and \textcolor{black}{$K$ is a constant that is defined in Eq.~\ref{pressure2}}. These two densities are meant to be continuous at the boundary, that
is, $\rho_{\rm s}= \rho_{\rm n}=\rho_{\rm c}$ at $r=R_{\rm c}$, although this boundary condition is independent of the coordinate system and spherical symmetry. These two phases could be continuously interpolated using $\rho = \rho_0 (1- m\Phi/k_{\rm B}T/n)^n$, where in $n \rightarrow 1/2$ we have the superfluid phase, and in $n \rightarrow \infty$ we have the isothermal phase. Alternatively, the real fluid can be thought of as being some superposition of the two fluid phases.

%Now that we have defined how DM behaves within the superfluid core, we must then prescribe the behaviour of the normal phase around the core. 

%First we impose the constraint that the density is continuous at the boundary of the superfluid, which means that our density profile should take the form,
%%
%\begin{equation}\label{densityprofile}
%\rho_{DM}(r) = \begin{cases}
%
%\rho_{\rm s}(r)   & \text{for  $ r\leq R_{\rm c}$} \\
%
%\rho_{\rm n}(r)   & \text{for $r\geq R_{\rm c}$} \\
%
%\end{cases}\,,
%\end{equation}
%%
%with the condition $\rho_{\rm s}= \rho_{\rm n}=\rho_{\rm c}$ at $r=R_{\rm c}$.

%Hence the pressure in the two phases are given respectively as 
%\begin{eqnarray}
%P & = &  {\rho^3 \over 12K^2},  ~\mbox{for~superfluid~density}~\rho =  2K (-2\Phi)^{1/2}\label{densitypotrelation} \\
%   & = &  \frac{k_{\rm B}T}{m} \rho ~\mbox{for~normal~fluid}~ \rho=\rho_0 \exp{-m \Phi(\vec{r}) \over k_{\rm B}T} \,.
%\end{eqnarray}
%
%We also simplify the isothermal model, such that we replace  $\rho = \rho_0 \exp{-m \Phi(r) \over k_{\rm B}T} $ with 
% $\rho = \rho_c (R_c/r)^2$.  The pressure and density are required to be continuous at $r=R_c$.
%
%The $\rho_{\rm c}$ in Eq.~\ref{normalphasedensity}, like the core radius, is yet to be determined (again described in Sec.~\ref{corecalc}). Therefore, once we determine $R_{\rm c}$ and $\rho_{\rm c}$, our normal phase component is fully defined.

\subsection{Matching the superfluid core to the normal halo}
\label{corecalc}

As mentioned above, we assumed a model where the superfluid phase abruptly transitions into the normal phase of the fluid. We therefore determined at which radius this transition occurs, \textcolor{black}{which defines the core radius $R_{\rm c}$}. To do this, we imposed the condition that the tangential pressure is continuous at the boundary between the core and normal-phase DM. Therefore, the determined radius at which this criterion is satisfied is our core radius $R_{\rm c}$. We outline this procedure step by step.

\begin{itemize}

\item {\large Step 1: Calculating the gravitational enclosed mass}

\vspace{0.2cm}
\noindent In order to perform the pressure-matching routing, we calculated the total enclosed gravitating mass (baryons + DM) as a function of radius for $r\geq R_c$:
\begin{align}\label{mass}
\begin{split}
M_{\rm grav}(r) &= M_{\rm c} + M_{\rm n}(r) + M_{\rm b}(r)\\
&= 4\pi \int^{R_{\rm c}}_{0} \rho_{\rm s}(r') r'^2 {\rm d}r' + 4\pi  \int^{r}_{R_{\rm c}} \rho_{\rm n}(r') r'^2 {\rm d}r' + M_{\rm b}(r) \\
&= 8\pi \sqrt{2}K \int^{R_{\rm c}}_{0}  \sqrt{-\Phi (r)} r'^2 {\rm d}r' + 4\pi R_{\rm c}^{2} \rho_{\rm c}\left[ r - R_{\rm c} \right] \\
& + M_{\rm b}(r) \,.
\end{split}
\end{align} 
%8 \pi r^{2} \sqrt{2} K \sqrt{-\Phi (r)} {\rm d}r\\ &+ \int^{r}_{R_{\rm c}} 4 \pi r^{2} \rho_{\rm c}\left(\frac{R_{\rm c}}{r'} \right)^{2} {\rm d}r\\ &+ M_{\rm b}(r)\,.\\
The first term comes from integrating the superfluid density~Eq.~\ref{superfluiddensity}, the second from integrating the normal-phase density~Eq.~\ref{normalphasedensity}, and the third is the baryonic mass. We therefore have the enclosed mass as a function of $R_{\rm c}$ and $\rho_{\rm c}$. \\

\item { \large Step 2: Solving the Poisson equation} 

\vspace{0.2cm}
\noindent Poisson's equation~(Eq.~\ref{Poissonbary2}) is solved numerically given an initial value of the gravitational potential $\Phi_{0}$ assuming the central gradient is zero. This yields the gravitational potential profile $\Phi(r)$, which can be substituted to obtain the superfluid density profile $\rho_{\rm s}(r) = 2K \sqrt{-2 \Phi(r)}$ as explained earlier. In our model, the superfluid density and gravitational potential profiles are only valid for $r \leq R_{\rm c}$. The next step is to calculate $R_c$. \\

\item {  \large Step 3: Calculate the core radius:} 

\vspace{0.2cm}
\noindent As mentioned, we assumed that the boundary between the normal and superfluid phase of DM is the point at which the tangential pressures are equal. Mathematically, this means the point at which
\begin{equation}\label{pressurebalance2}
\frac{\rho_{\rm c}^{3} }{12 K^{2}} =\int^{\infty}_{R_{\rm c}} \frac{\rho_{\rm n}(r) G_{\rm N} M_{\rm grav}(r)}{r^{2}}{\rm d}r = R_{\rm c}^{2} \rho_{c} \int^{\infty}_{R_{\rm c}} \frac{G_{\rm N} M_{\rm grav}(r)}{r^{4}}{\rm d}r\,.
\end{equation}
The left-hand side comes from the polytropic equation of state~Eq.~\ref{pressure2} describing the superfluid, with $\rho_{\rm c} = \rho_{\rm s}(R_{\rm c})$. The right-hand side follows from the hydrostatic equilibrium equation of the normal-phase density, where we have substituted~Eq.~\ref{normalphasedensity} for $\rho_{\rm n}(r)$. Furthermore, $M_{\rm grav}(r)$ refers to the total enclosed mass (baryonic + DM) at radius r.

In order to solve Eq.~\ref{pressurebalance2} for the core radius $R_{\rm c}$, we must write the parameters $\rho_{\rm c}$ and $M_{\rm grav}(r)$ as functions of $R_{\rm c}$.
The former is obtained by evaluating Eq.~\ref{superfluiddensity} at $R_c$:
\begin{equation}\label{coredensity}
\rho_{\rm c} = \rho_{\rm s}(R_{\rm c}) = 2 K \sqrt{-2 \Phi(R_{\rm c})}\,.
\end{equation}
For the enclosed mass $M_{\rm grav}(r)$, we note that the lower limits of the integral on the right-hand side of Eq.~\ref{pressurebalance2} are greater than $R_{\rm c}$, hence we can use our expression Eq.~\eqref{mass}, which is valid for $r \geq R_c$. Therefore by numerically solving Eq.~\ref{pressurebalance2} using the derived expressions, for the mass (Eq.~\ref{mass}) and core density (Eq.~\ref{coredensity}) we can determine a value for the core radius, $R_{\rm c}$. From the calculated core radius, we can find the numerical value for the core density via Eq.~\ref{coredensity}.

\end{itemize}

\subsection{Determining the virial radius}

In the previous section we began by defining a central potential $\Phi_{0}$ and from this we determined the core radius as well as the core density. We therefore have at our disposal all the necessary components to determine the density of the DM fluid both inside and outside the core. 

The next stage is to determine the radius at which to truncate the DM halo. In the $\Lambda$CDM paradigm, this is commonly defined for clusters as $r_{200}$, the radius at which the average density falls to 200 times the critical density of the universe at a given redshift \citep{sample}, that is, the virial density. { We adopt this convention in our model. The mass $M(R_{\rm vir})$ is then determined by evaluating Eq.~\ref{mass} such that the above convention is respected at $R_{\rm vir}$.} 
%First, we should remind ourselves of the critical density, defined as
%
%\begin{equation}
%\rho_{\rm crit}(z_{\rm vir}) = \frac{3 H^{2}(z_{\rm vir})}{8\pi G_{\rm N}} \,,
%\end{equation} 
%
%where $H(z)$ is the Hubble parameter determined from the background evolution, and $z_{\rm vir}$ is the %\textcolor{black}{redshift at which the cluster becomes virialized}. The virial radius is then determined by %solving
%
%\begin{equation}
%\frac{3 M(R_{\rm vir})}{4\pi R_{\rm vir}^{3}} = 200 \rho_{\rm crit}(z_{\rm vir})\,.
%\end{equation}
%

\subsection{Making mass the free parameter}\label{massparam}

We have now determined a simplified way to model a spherical system in the superfluid framework. However, this was achieved by initially solving Poisson's equation, which in turn required the input of the central potential $\Phi_{0}$ as a free parameter. A more ideal set-up would be to determine all relevant quantities, $R_{\rm c}$, $\rho_{\rm c}$, $R_{\rm vir}$ , and $\Phi_{0}$ from a known total mass $M_{\rm vir}$. This is possible with the aid of interpolation. The above procedure for a discrete set of $n$ initial central potential values might be carried out, for example, $\lbrace\Phi_{01}, \Phi_{02}...\Phi_{0n}\rbrace$. From this set an array of core radii is built, $\lbrace R_{c01}, R_{c02}...R_{c0n}\rbrace$, as well as core densities $\lbrace\rho_{c01}, \rho_{c02}...\rho_{c0n}\rbrace$, virial radii $\lbrace R_{{\rm vir}01}, R_{{\rm vir}02}...R_{{\rm vir}0n}\rbrace,$ and virial masses $\lbrace M_{{\rm vir}01}, M_{{\rm vir}02}...M_{{\rm vir}0n}\rbrace$. Plots of $\Phi_{0}$ {\it vs} $M_{\rm vir}$, $R_{\rm c}$ {\it vs} $M_{\rm vir}$, $\rho_{\rm c}$ {\it vs} $M_{\rm vir}$, and $R_{\rm vir}$ {\it vs} $M_{\rm vir}$ are then possible. Then, interpolation procedures can be implemented to make continuous functions of virial mass as functions of core radius, core density, central potential, and virial radius. This allows picking a virial mass and easily determining the required parameters that describe the fluid. 

\section{A worked example without baryons}\label{toy}

\begin{figure*}
\centering
\begin{tabular}{cc}
\includegraphics[scale=0.7]{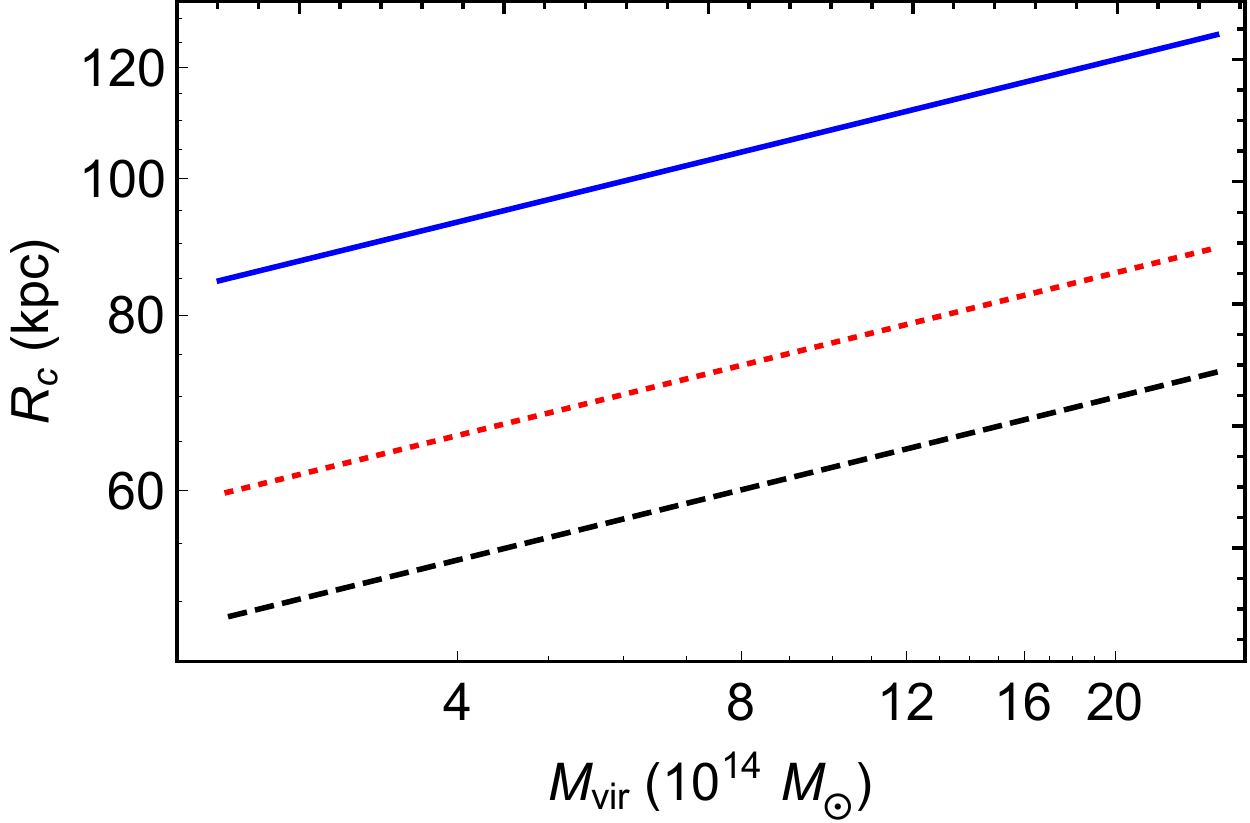} & \includegraphics[scale=0.7]{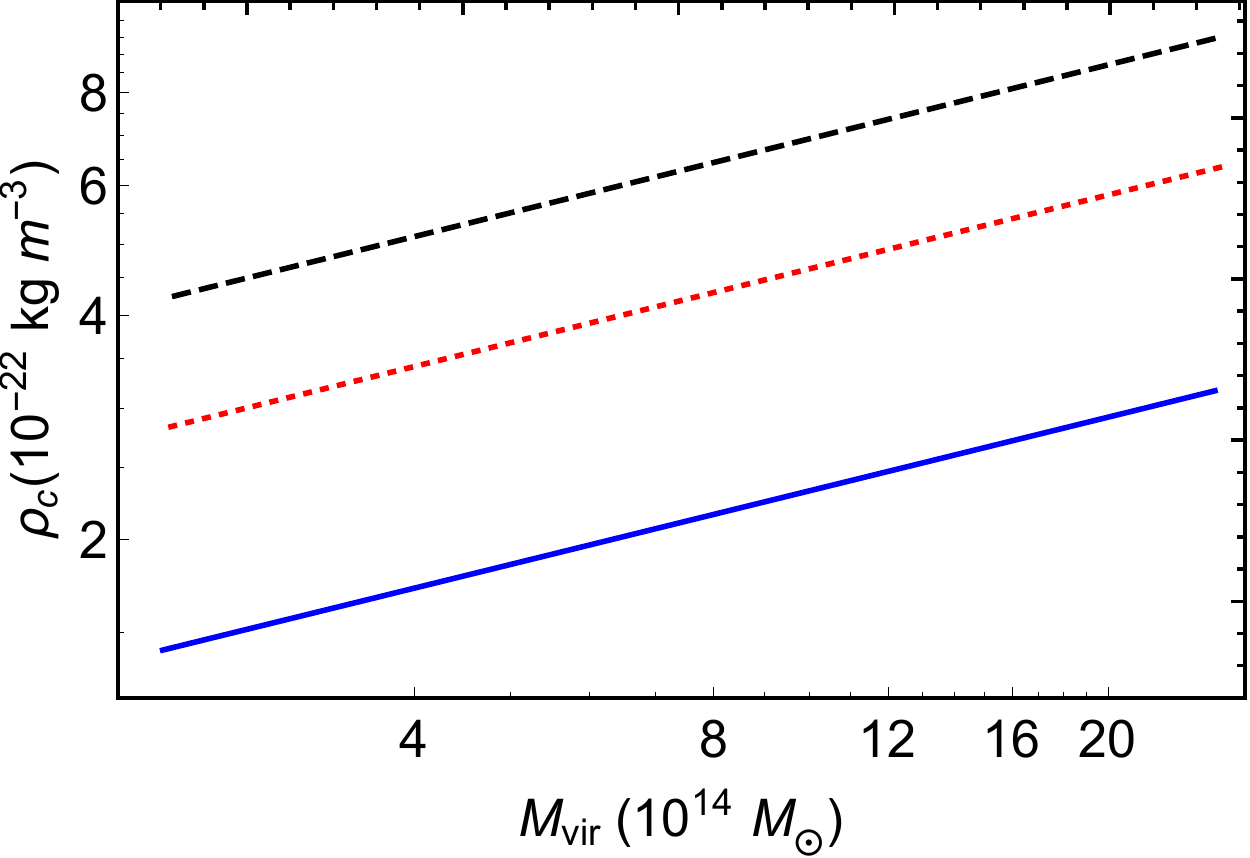}\\
\includegraphics[scale=0.7]{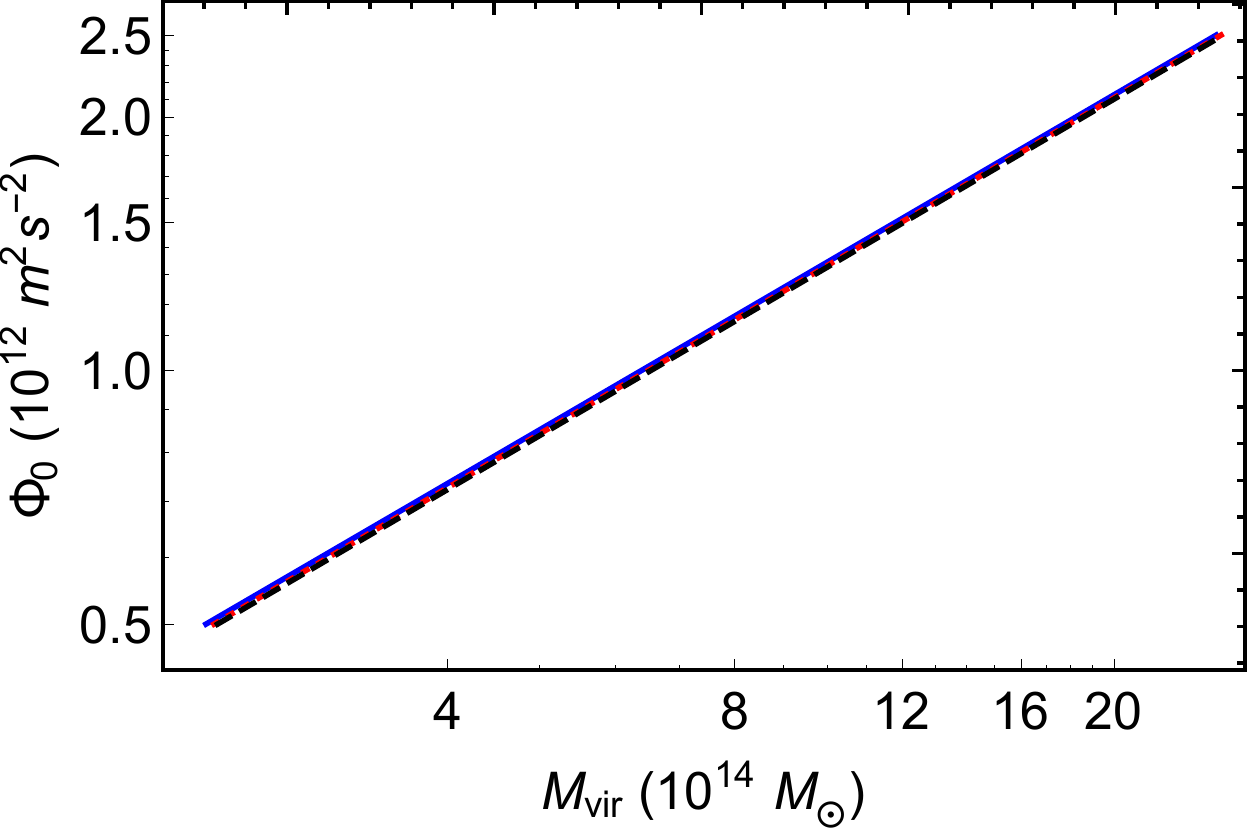}&
\includegraphics[scale=0.7]{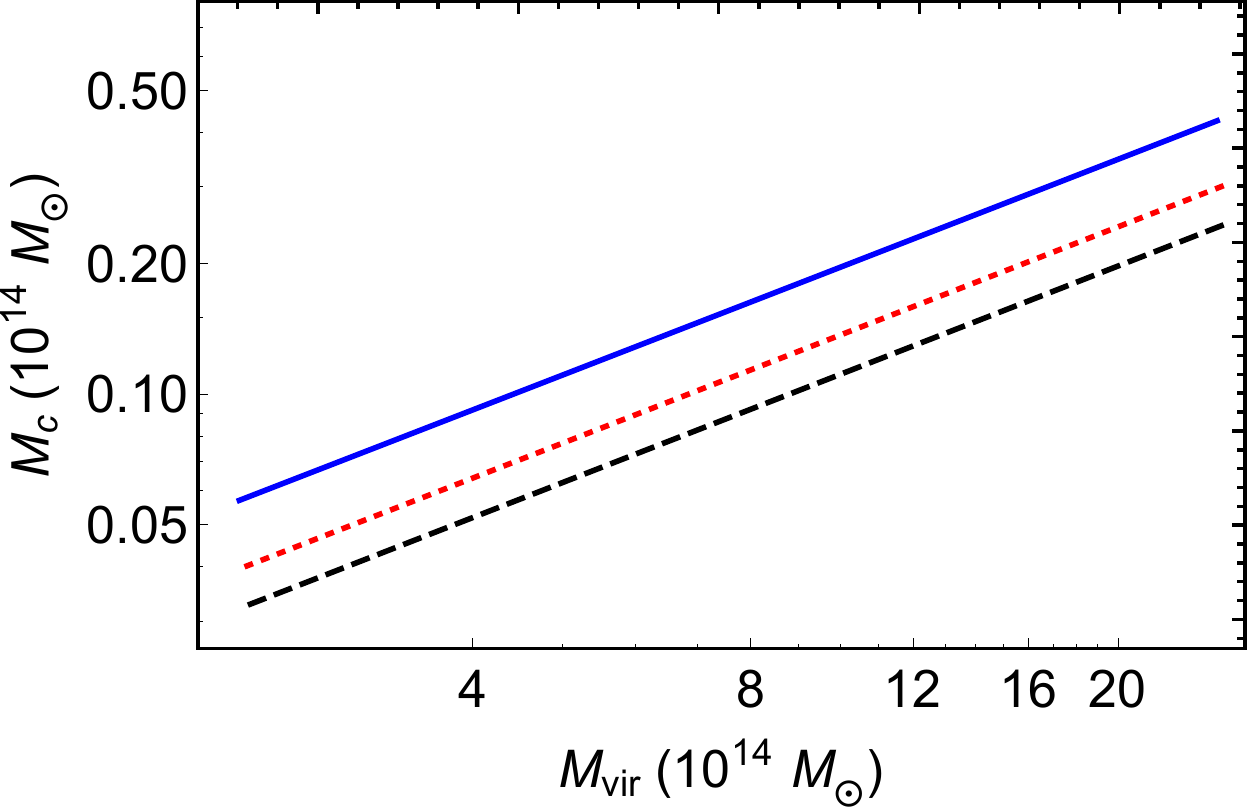}\\
\end{tabular}
\caption{Toy model of a DM-only system. We highlight the dependence of the superfluid parameters on the choice of virial mass for a system that only includes DM. We stress that this result does not consider baryons, and a more complete analysis should be performed, which we provide for the cluster sample. If we were to include baryons, we would expect to see a slight decrease in core radius for our clusters. This difference would be enhanced in galaxies. Top left: core radius vs virial mass. Top right: density at the core radius vs virial mass. Bottom left: central potential vs virial mass. Bottom right: superfluid core mass vs virial mass. The three lines in each plot represent different choices of $\Lambda m^{3}$. The blue solid line has $\Lambda m^{3} = 0.1$~$\times~ 10^{-3}$ eV$^{4}/c^{8}$, the dotted red line has  $\Lambda m^{3} = 0.2$~$\times~ 10^{-3}$ eV$^{4}/c^{8}$ , and the dashed black line has $\Lambda m^{3} = 0.3$~$\times~ 10^{-3}$ eV$^{4}/c^{8}$.}
\label{interpplotsdmonly}
\end{figure*}

We first illustrate here our procedure for a DM-only galaxy cluster, that is, without baryons, and with an isothermal normal phase. Figure~\ref{interpplotsdmonly} shows the interpolation functions for the core radius $R_{\rm c}$, density at core radius $\rho_{\rm c}$, central potential $\Phi_0$ , and superfluid core mass $M_{\rm c}$ as functions of the virial mass $M_{\rm vir}$, following the procedure outlined in Sect.~\ref{massparam}. These allow us to determine the correct parameters for a given total virial mass. In this example we chose a total mass of $10^{15} M_{\odot}$ at redshift $z = 0$. 

Figure~\ref{interpplotsdmonly} shows that $R_{\rm c}$, $\rho_{\rm c}$ and $\Phi_0$ all scale with virial mass by power-law profiles. We can try to understand this by analysing the model at the core radius. Firstly, ignoring baryons, Eq.~\ref{pressurebalance2} can by simplified to
\begin{equation}\label{interpanalysis1}
\frac{\rho_{\rm c}^{2}}{12 K^{2}} = G_{\rm N} \left( \frac{M_{\rm c}}{R_{\rm c}}- 2\pi R_{\rm c}^{2}\rho_{\rm c} \right)\,.
\end{equation}
To understand the origin of the power laws, we did not perform a rigorous calculation, but instead used approximations and dimensional analysis to reason that the numerical results are reasonable. As mass is proportional to volume and density, we can make the approximate proportional relation, $M_{\rm c} \propto \rho_{\rm c} R_{\rm c}^{3}$. Therefore Eq.~\ref{interpanalysis1} can be interpreted as $\rho_{\rm c} \propto R_{\rm c}^{2}$. We can also make the approximation that the velocity $v$ is proportional to the virial velocity $v_{\rm vir}$, which in turn we assume is proportional to the sound speed $c_{s}$. From our knowledge of the sound speed from Sect.~\ref{DMsuperfluidphase} we know that $c_{s} \propto \sqrt{\Phi}$. From our Poisson \textcolor{black}{equation} we also know $\sqrt{\Phi}\propto \rho_{\rm c}$. Therefore $v_{\rm vir} \propto \rho_{\rm c}$. The virial radius is related to the virial mass as $R_{\rm vir}^{3} \propto M_{\rm vir}$. Thus from the relation for circular velocity, we obtain $v_{\rm vir} \propto M_{\rm vir}^{1/3}$. Collecting all these approximations, we obtain the following scaling relations:

\begin{equation}
\begin{split}
\rho_{\rm c} &\propto M_{\rm vir}^{1/3} \\
R_{\rm c} &\propto M_{\rm vir}^{1/6} \\
\Phi_{0} &\propto M_{\rm vir}^{2/3} \\
M_{\rm c} &\propto M_{\rm vir}^{5/6}\,,\\
\end{split}
\end{equation}

\noindent which are in agreement with our numerical results. Although very informal, this analysis allows us to  make an educated
guess as to how the quantities might scale with each other. These scaling relations will not hold in the case when baryons are included, however.

\begin{figure}
\centering
\begin{tabular}{c}
\includegraphics[scale=0.7]{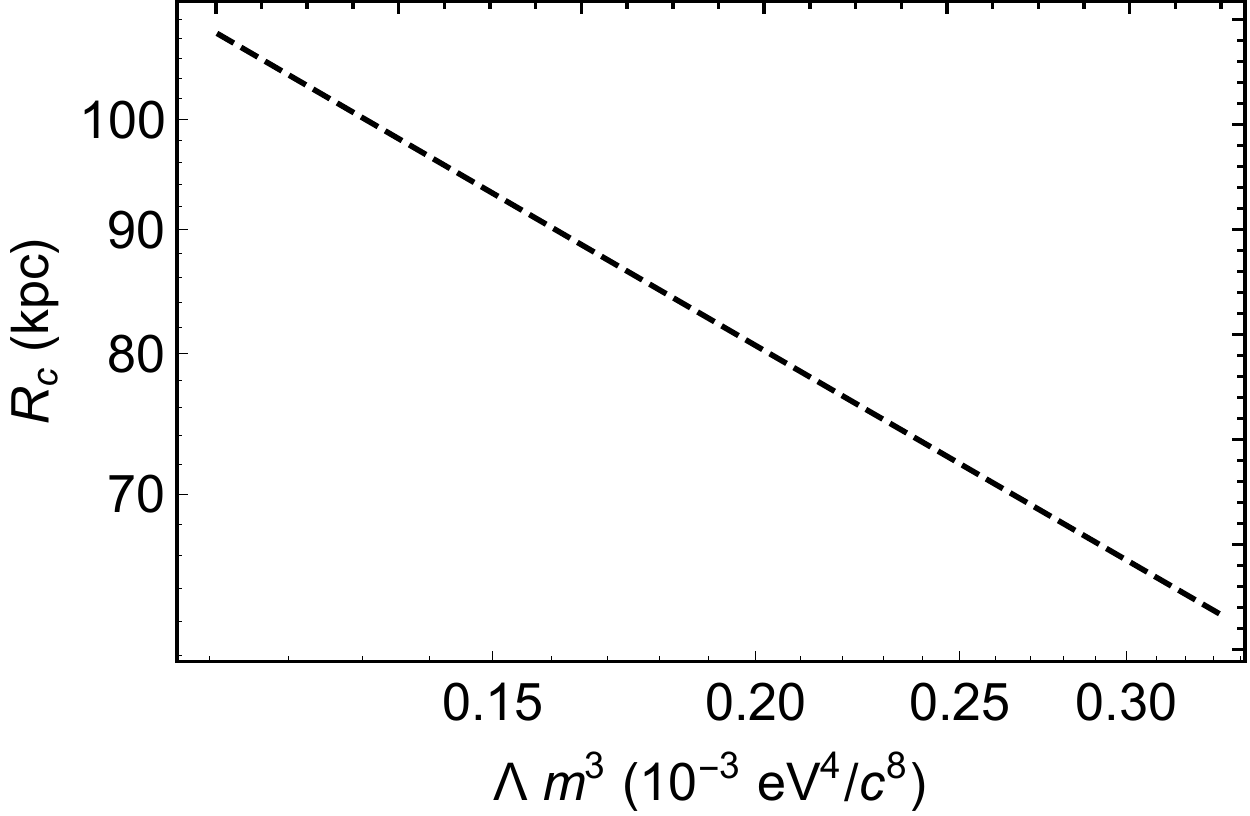}\\ \includegraphics[scale=0.7]{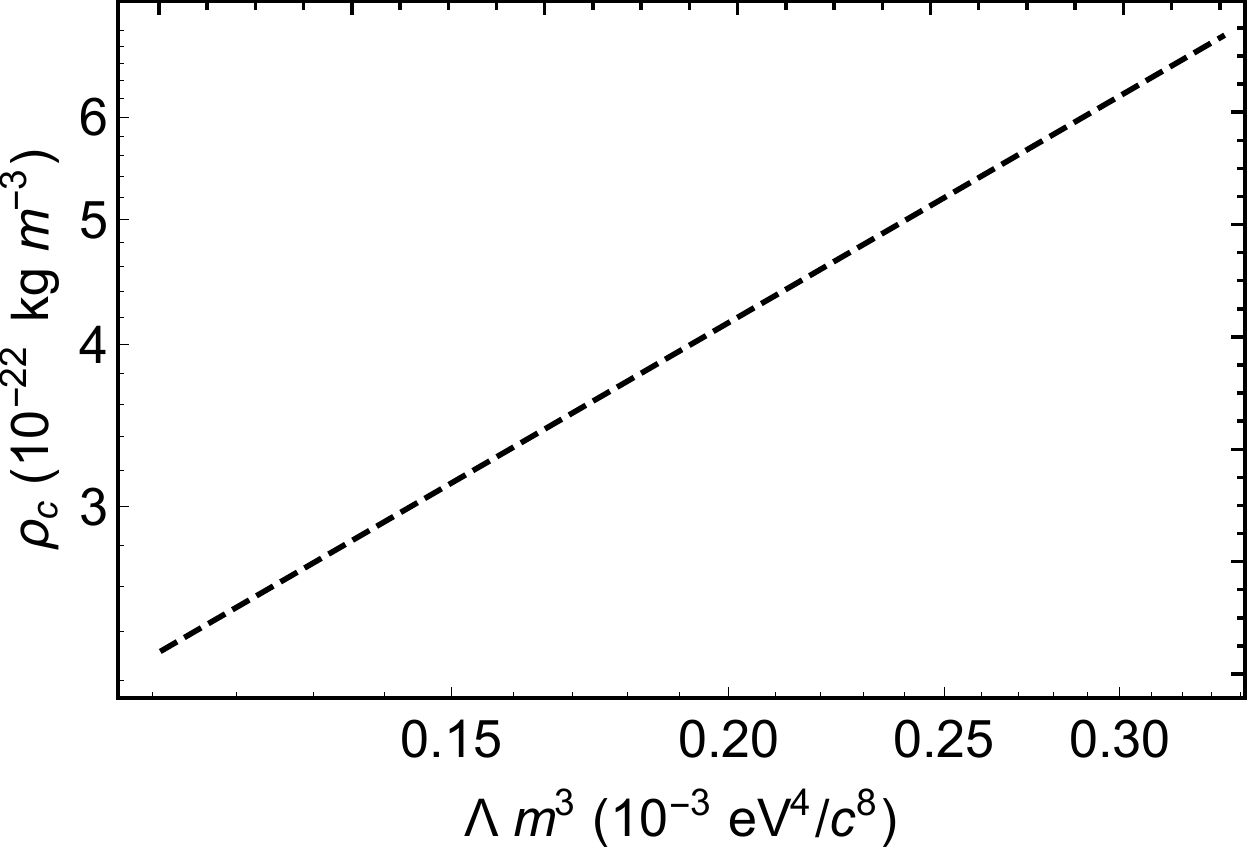}\\
\includegraphics[scale=0.7]{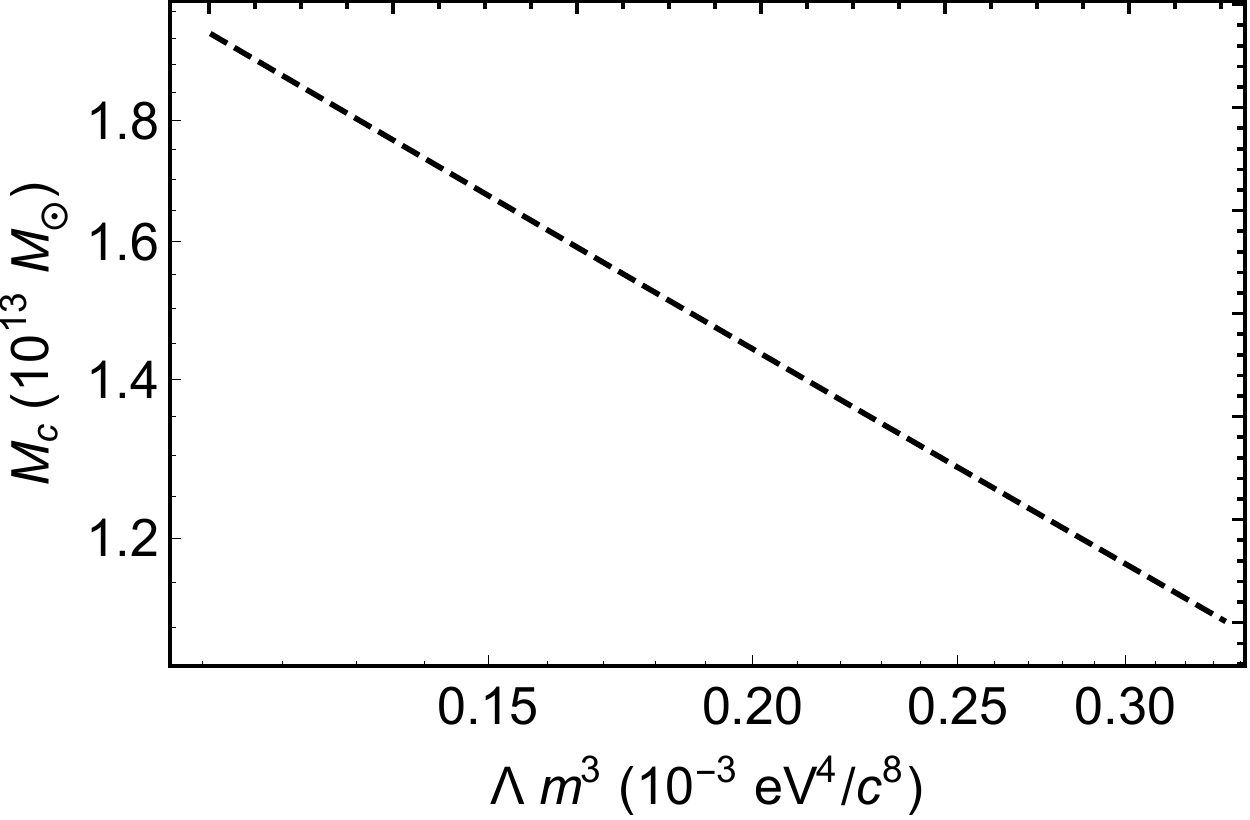}\\
\end{tabular}
\caption{\textcolor{black}{Change in core radius (top panel), density at core radius (middle panel), and core mass (bottom panel) as a result of different values of $\Lambda m^{3}$. This is a visual representation of the relationships \textcolor{black}{listed} in Eq. ~\ref{Lm3dependence}.}}
\label{Lm3NoDMPlot}
\end{figure}

We also show in Fig.~\ref{Lm3NoDMPlot} how varying the $\Lambda m^{3}$ parameter affects the results. When we increase $\Lambda m^{3}$ , we find a smaller $R_{\rm c}$, larger $\rho_{\rm c}$ , and lower $M_{\rm c}$, at the same virial mass. On the other hand, $\Phi_0$ remains unchanged. Again, this dependence can be understood on dimensional grounds. 
Specifically, the $\Lambda m^{3}$ scaling relations are 
\begin{equation}\label{Lm3dependence}
\begin{split}
\rho_{\rm c} &\propto \Lambda m^{3}\\
R_{\rm c} &\propto \left(\Lambda m^{3}\right)^{-1/2}\\
\Phi_{0} & \propto \left(\Lambda m^{3}\right)^{0}\\
M_{\rm c} &\propto \left(\Lambda m^{3}\right)^{-1/2}\,.\\
\end{split}
\end{equation}

These scaling relations can be understood by looking at the
dimensionless form of Eqs.~\ref{Poissonbary2} and \ref{pressurebalance2}.  Omitted here for conciseness,
it can be shown that a dimensionless core radius parameter arises
from the Poisson equation,

\begin{equation}
x_c \equiv \frac{\rho_c}{32\pi G K^2 R_c^2}\,.
\end{equation} 

\noindent This can also be understood from straightforward dimensional analysis;
note that $K$ has the dimension of density per velocity.  
The same $x_c$ arises in the pressure-balancing condition (Eq. \ref{pressurebalance2}), which 
then fixes $x_c$ as a numerical constant for all halos.  
As $x_c$ is constant, we gain information that $\rho_c/R_c^2$ scale as $K^2$.  

Under the condition of a small core, that is\textcolor{black}{, $R_{\rm c}  \ll R_{\rm vir}$, hence $M_{\rm c} \ll M_{\rm vir}$}, we obtain
from Eq.~\ref{mass} the relation
\begin{equation} 
{M_{\rm vir} \over 4\pi R_{\rm vir}}  = \rho_{\rm c} R_{\rm c}^2 \,.
\end{equation}
We therefore have two relations, which, together with $R_{\rm vir}$ $\sim v_{\rm vir}$ $\sim M_{\rm vir}^{1/3}$, fix the scaling relations,
$\rho_{\rm c} \sim K v_{\rm vir}$,  $R_{\rm c} \sim \sqrt{v_{\rm vir}/K}/G $.  The potential $\Phi_{\rm 0} \sim G \rho_{\rm c} R_{\rm c}^2 \sim v_{\rm vir}^2$
is independent of $K$. We can finally use Eq.~\ref{interpanalysis1} to find the $M_{\rm c}$ dependence.

These are in good agreement with the numerical results. \textcolor{black}{To highlight this, \textcolor{black}{in} Figure ~\ref{Lm3NoDMPlot} we plot the dependence of $R_{c}$, $\rho_{c}$ , and $M_{c}$ on $\Lambda m^{3}$ for a fixed virial mass. We do not include the plot of $\Phi_{0}$ as it is not affected by changing $\Lambda m^{3}$, which we checked numerically. Figure ~\ref{Lm3NoDMPlot} follows the relationships shown in \textcolor{black}{Eq.} \ref{Lm3dependence}.}

As we have the relationship of the superfluid parameters with the virial mass and $\Lambda m^{3}$, we can determine the correctly normalised scaling relations for the core radius, density at core radius, central potential, and the core mass. These are
\begin{equation}
R_{\rm c} \approx 36.2 \; \left(\frac{M_{\rm vir}}{10^{15}M_\odot}\right)^{1/6}\left(\frac{\Lambda m^3}{10^{-3} {\rm  eV}^4/c^{8}}\right)^{-1/2}~{\rm kpc}\,,
\end{equation}

\begin{equation}
\rho_{\rm c} \approx 2.1\times 10^{-21}\; \left(\frac{M_{\rm vir}}{10^{15}M_\odot}\right)^{1/3}\left(\frac{\Lambda m^3}{10^{-3} {\rm  eV}^4/c^{8}}\right)~{\rm kg~ m^{-3}}\,,
\end{equation}

\begin{equation}
M_{\rm c} \approx 6.5 \times 10^{12}\; \left(\frac{M_{\rm vir}}{10^{15}M_\odot}\right)^{5/6}\left(\frac{\Lambda m^3}{10^{-3} {\rm  eV}^4/c^{8}}\right)^{-1/2}~{\rm M_\odot}\,,
\end{equation}

\begin{equation}
\Phi_{\rm 0} \approx -1.3 \times 10^{12}\; \left(\frac{M_{\rm vir}}{10^{15}M_\odot}\right)^{2/3}~{\rm m^{2}s^{-2}}\,.
\end{equation}

\noindent \textcolor{black}{These only rigorously apply for the situation where the baryon contribution is omitted and the normal phase is isothermal.}

\begin{figure}
\centering
\includegraphics[scale=0.7]{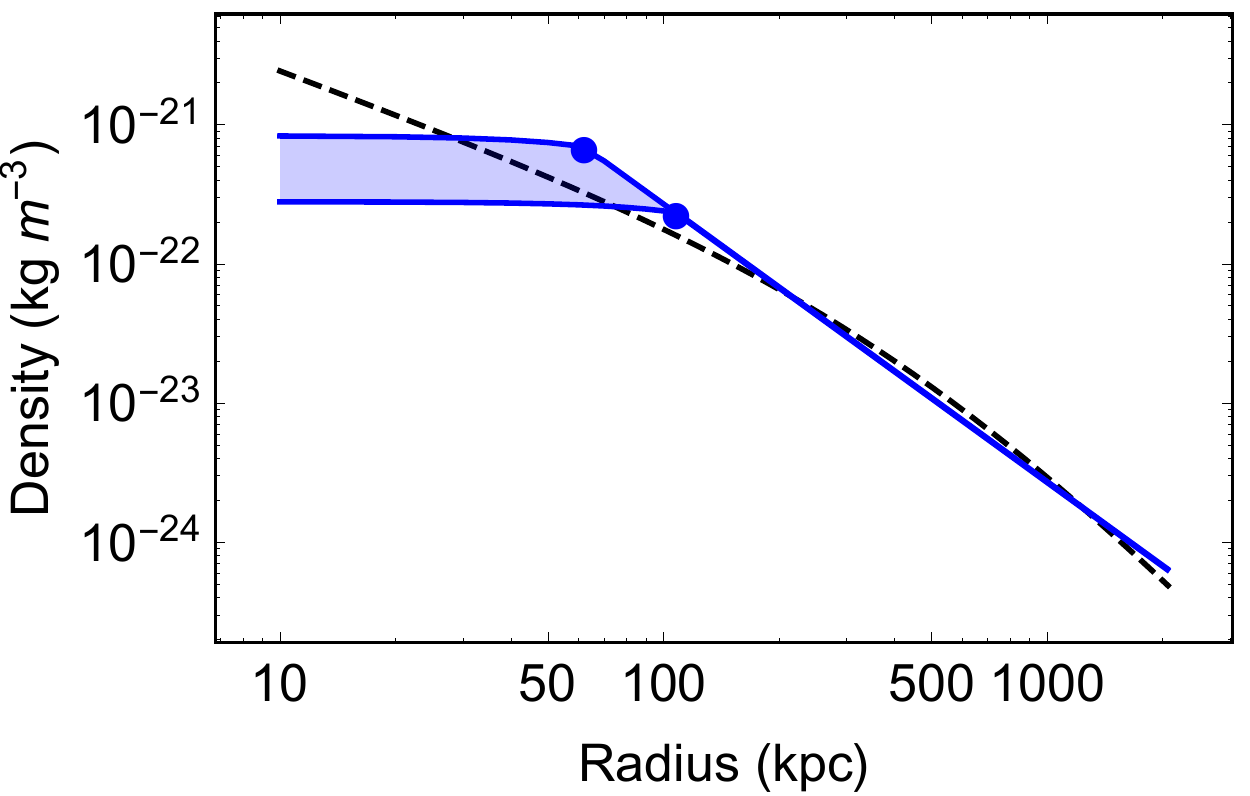}
\caption{Shows the DM only density profile for our SfDM toy model with $\Lambda m^{3} = 0.1-0.3$~$\times~ 10^{-3}$ eV$^{4}/c^{8}$ (blue shaded) and a $c_{200}=4$ $\Lambda$CDM NFW comparison profile (dashed black line) each with a virial mass of $10^{15} M_{\odot}$. Blue circles show the radii of the superfluid core for different  $\Lambda m^{3}$ choices \textcolor{black}{(the upper line represents a larger choice of $\Lambda m^{3}$, which results in a higher density and smaller core radius)}. Inside the core, the density is approximately constant, outside the core the density follows a strict isothermal $1/r^{2}$ power law. {The inclusion of baryons in this model will not affect the general features too much}, but will shrink the core radius slightly. The core radius can be increased by using a smaller value of $\Lambda m^{3}$, which we show in the following sections of the paper.}
\label{denprofile}
\end{figure}

Figure~\ref{denprofile} shows the range of DM density profiles in our model for different parameter choices (blue shaded region), together with an NFW profile of concentration $c_{200} = 4$ (black dashed line)  for comparison. The virial mass is the same in the two cases, $M_{\rm vir} = 10^{15}M_\odot$. \textcolor{black}{The blue dots highlight the core radius for different choices of $\Lambda m^{3}$.} We see that the superfluid density is relatively constant within the core, as expected. Outside the core, the density profile transitions to the normal-phase, isothermal profile with $\rho \sim 1/r^2$. \textcolor{black}{We can see from Fig.~\ref{denprofile} that the core radius we calculate is approximately 50-100 kpc for different parameter choices.} 

We truncated the density to zero at the virial radius. If left untruncated, the enclosed mass would keep growing linearly, which is unphysical. Although the choice of the virial radius is somewhat arbitrary, as mentioned earlier, we chose $r_{200}$ because this is commonly adopted in the literature.

\begin{figure}
\centering
\begin{tabular}{c}
\includegraphics[scale=0.7]{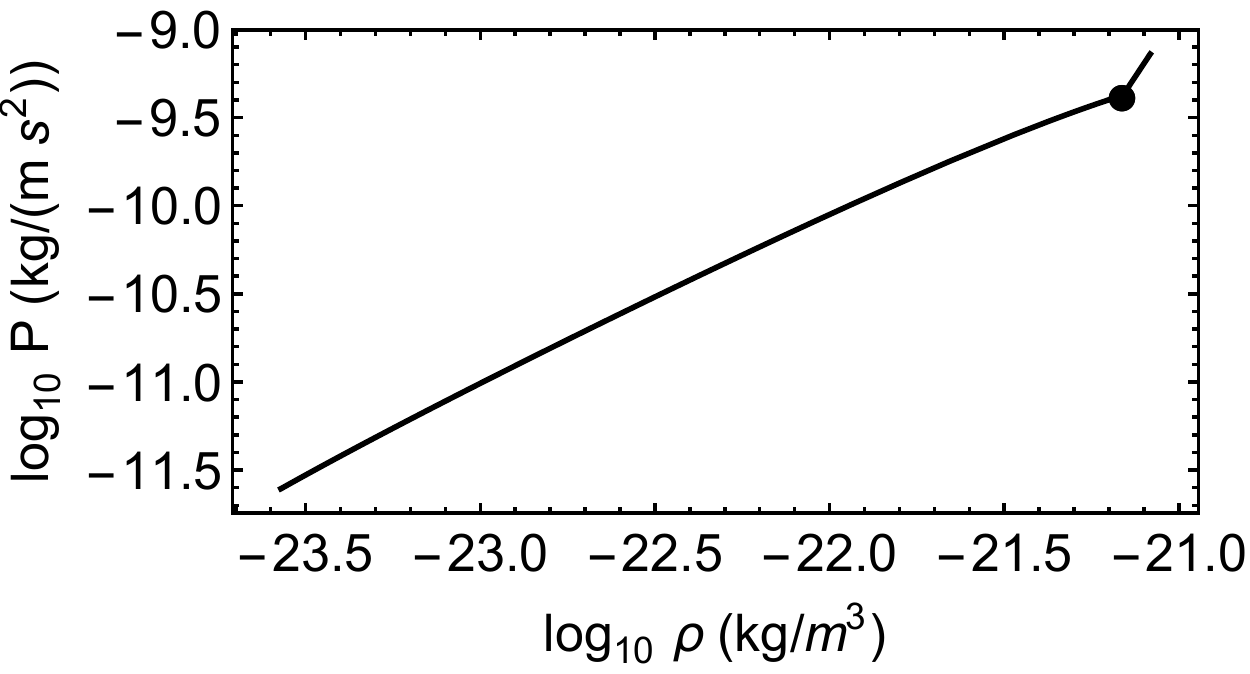}\\
\includegraphics[scale=0.6]{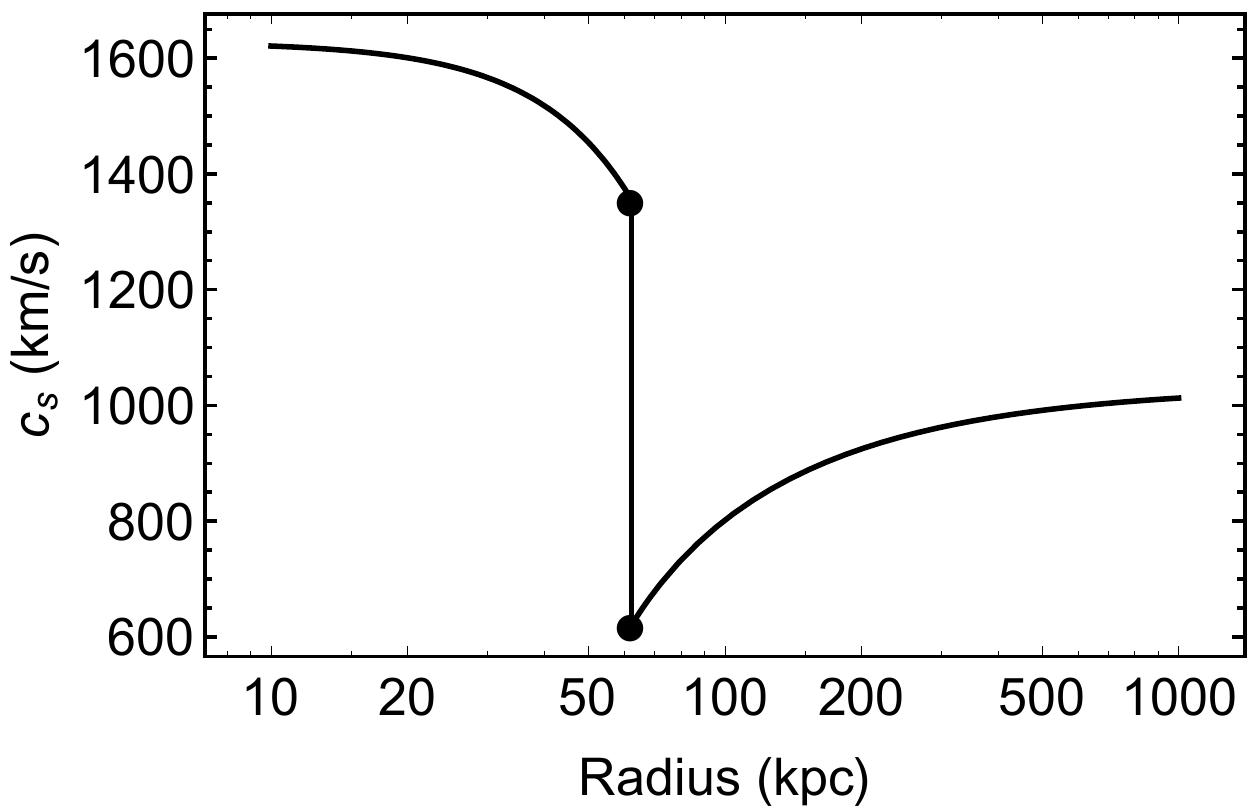}
\end{tabular}
\caption{\textcolor{black}{Top panel: pressure vs density for a DM-only model. Both the pressure and density are continuous, as per our phase transition requirements. At the core radius (black circle), the gradient jumps because of the discontinuity in \textcolor{black}{the equation of state at the transition between the superfluid and normal phase of DM. The core radius is located very close to the rightmost part of the plot as the density is approximately constant within the core.
Bottom panel: sound speed of DM (\textcolor{black}{$c_{\rm s}^{2} = \rm {d}P/\rm {d}\rho$}). \textcolor{black}{The jump at the core radius (vertical line) results from the discontinuous equation of state at the boundary}. This might be a consequence of our simplified model of the phase transition. Addressing this issue is best left for future work. Both panels have $\Lambda m^{3} = 0.3$~$\times~ 10^{-3}$ eV$^{4}/c^{8}$}} }
\label{Pvsrho}
\end{figure}

\textcolor{black}{To highlight the continuity of density and pressure, we show in Fig.~\ref{Pvsrho} pressure vs density (top panel) and sound speed vs radius (bottom panel). Both the pressure and density are continuous, and thus we have satisfied our phase-matching criteria. However, we note a sudden jump in gradient at the core radius. This is due to our requirement that density and pressure must be continuous, but not a continuity of \textcolor{black}{the equation of state}. This results in a discontinuous sound speed. This is perhaps a limitation of our model and our simplified transition between the superfluid and normal phase of DM. We do not address this matter here. }

{ We previously mentioned that in reality, the superfluid core would continue radially until its density dropped to zero. There should  be a transition zone within which the superfluid and normal phases co-exist.} Introducing this feature would make the model much more complex, as understanding how to define the normal-phase profile would become more challenging.

Simplicity is the actual reason for adopting our two-phase model with the core radius occurring when the density and tangential pressure are continuous. The result of this is a superfluid phase that has a much smaller radial extent than the true model, meaning that the core radii values we quote here are underestimated. 

\begin{figure}
\centering
\includegraphics[scale=0.7]{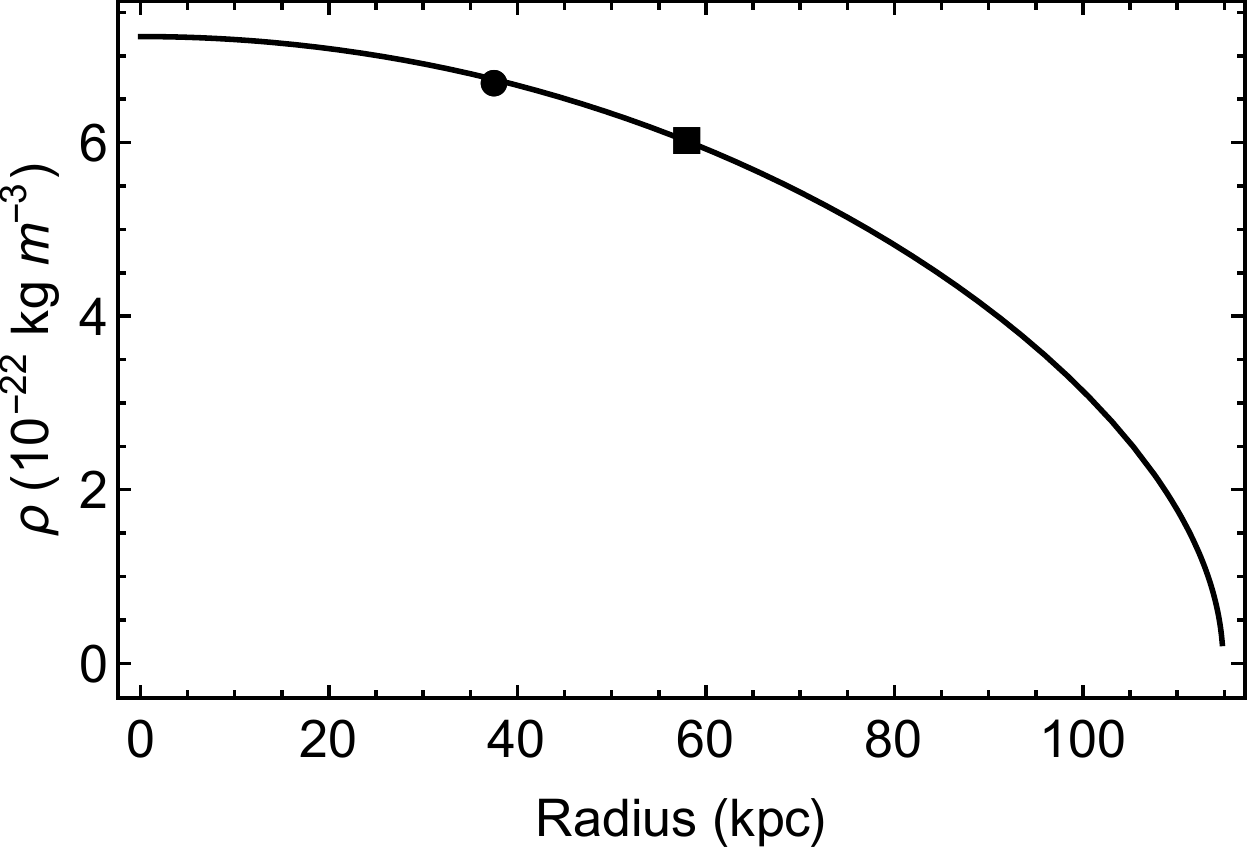}
\caption{{ Superfluid-only} density profile for a { DM-only} system with central gravitational potential $\Phi_{0} = -10^{12}$ m$^{2}$s$^{-2}$ and $\Lambda m^{3} = 0.3$~$\times~ 10^{-3}$ eV$^{4}/c^{8}$. { The ``true" core radius at which the effect of phonons disappears completely is where the superfluid density drops to zero. However, in the modelling presented in this paper, DM is considered to be entirely in the normal-phase (non-superfluid) beyond the radius at which the tangential pressure is matched (what we call $R_{c}$).} This matching radius is illustrated by a square for the isothermal case, and a circle shows the NFW normal-phase case. We show this to emphasise that our method truncates any superfluid phonon effects at a smaller radius than the true radius of the superfluid matter. This is a feature of our simplified model, and more complex models will have to be constructed in the future to fully understand the superfluid paradigm.}
\label{Rcprofile}
\end{figure}

{ To highlight this, we plot the superfluid-only density profile for a DM-only model (Fig.~\ref{Rcprofile}) without considering the normal phase. The ``true" core radius is the point where $\rho_{s} = 0$. However, the normal phase should already be dominating at this stage. There is thus a whole ``transition zone" in which the relative contribution of the normal phase would rise while the superfluid density sharply decreases. }

We indicate the core radius we adopt for this work in the plot (black circle for an NFW-like normal phase and a black square for an isothermal normal phase). The ``true" core radius at which $\rho_{s} = 0$ is approximately twice the value we determine from our pressure-matching method for the isothermal case and approximately three times the value we determine for our NFW case. 

We can therefore conclude that we expect, in reality, superfluid phonon behaviour to occur at larger radii than we present here. Although the NFW case predicts the density and pressure to match at a smaller radius than the isothermal case, the ``true" physical core radius would be unaffected. For a future more realistic model, we need to investigate to which degree introducing a mixing phase would affect the galaxy clusters. This is beyond the scope of the present exploratory work.

\section{Application to galaxy cluster mass profiles}\label{dynamical}

After outlining the procedure for calculating the DM density profile, we turn to a sample of galaxy clusters to check whether the superfluid paradigm provides a good description of the dynamical mass deduced from the temperature profile of X-ray emitting gas in clusters, assuming hydrostatic equlibrium. 

{ To determine how well the superfluid paradigm describes real galaxy clusters, we therefore compare the expected DM profile, determined from our aforementioned simplified superfluid modelling, to the actual mass of galaxy clusters deduced from the temperature profile of the X-ray emitting gas.} This amounts to a comparison of the enclosed mass derived from integrating the density profile of our model and the mass derived from hydrostatic equilibrium arguments. We have discussed how the enclosed mass derived from the density profile would be calculated (see Eq.~\ref{mass}). We now briefly discuss how we derived the dynamical mass, that
is, the mass derived from the X-ray gas emission profile. 

Our treatment makes a number of simplifying assumptions \textcolor{black}{that are common when modelling the mass profile of galaxy clusters}. Firstly, the gas in the cluster is treated as an ideal gas, and secondly, the cluster is assumed to be in hydrostatic equilibrium. By assuming this, the following equation can be constructed linking the dynamical (or total) mass $M_{\rm dyn}(r)$ of the system within a radius \textcolor{black}{$r$},
\begin{equation}
 -\rho_{\rm g}(r) \frac{G_{\rm N} M_{\rm dyn}(r)}{r^{2}} = \frac{{\rm d}P_{\rm g}(r)}{{\rm d}r}\,.
\label{hydrostaticequi}
\end{equation}
Here, $\rho_{\rm g}$ is the gas density, $P_{\rm g}$ its pressure (linked to density and temperature through the ideal gas equation of state), and $M_{\rm dyn}(r)$ is the dynamical mass.

The dynamical mass is defined as the required amount of mass such that the gas temperature profile in hydrostatic equilibrium can be explained by Newtonian gravity. This mass can be compared to $M_{\rm grav}$ from Eq.~\ref{mass}. In $\Lambda$CDM and our superfluid cluster model the gravitational mass should be equivalent to the baryons + DM. In MOND the gravitational mass would be baryons + phantom dark matter. A phantom component in the superfluid paradigm arises from the phonons, which should be negligible in clusters. 

We have selected a sample of four galaxy clusters from \citet{sample}. At this stage, we only tested the model on a few clusters to gain an understanding of the current model predictions. We expect similar results for the other clusters in the~\cite{sample} sample and therefore consider it sufficient to leave extending the sample size for future work when the model is developed further. In this work, analytic profiles were prescribed for both the gas density and temperature. Fitting routines were then run to determine the parameters for each model. This allows the dynamical mass to be calculated analytically. Firstly, the gas emission profile and temperature profile are defined in Eqs.~3 and~6 of \cite{sample}, respectively.
The emission profile described in \cite{sample} is essentially a superposition of two $\beta$ density profiles that are commonly used when describing galaxy gas density. \citet{sample} added in the extra components to account for the steepening brightness at $r \approx 0.3 r_{200}$ and to impose a cuspy core to better match \textcolor{black}{observations}. More explicit details of this profile can be found in \citet{sample}.

In addition to the gas component, we also need to take the brightest cluster galaxy (BCG) into account, which is the large central galaxy component of the cluster. We modelled the baryonic component of the BCG using a Hernquist profile \citep{hernquistprofile},
\begin{equation}
\rho_{\rm BCG} (r) = \frac{M h}{2\pi r \left(  r + h \right)^{3}}\,,
\end{equation}
\noindent where $M$ is the mass of the BCG and \textcolor{black}{$h$} is the BCG scale radius. The Hernquist profile is an appropriate choice for modelling the BCG baryonic mass content as it provides a constant-mass profile at large radii, resulting in a well-defined size or radius of the BCG. We adopted the BCG mass from \cite{BCG2}, where the baryonic mass is approximated $1.14\times 10^{12} ~ {\rm M_{\odot}}$ for each cluster. In their model, they used a Jaffe profile with a scale length of $\approx 30~{\rm kpc}$. The Jaffe model has a slightly steeper density profile in the centre than the Hernquist model, which we mimicked by using a Hernquist scale radius of $10~{\rm kpc}$. Therefore the baryonic mass that we used when deriving the DM profile is the sum of the BCG and the gas masses. This is a crude approximation, and further work into understanding how large a BCG is required to fit the data is required. \textcolor{black}{We note that there does not seem to be a consistent way to prescribe the BCG mass in galaxy clusters. Current techniques include assuming a constant mass-to-light ratio \citep[e.g.][]{angus20081} or imposing a dependence on the total mass \citep[e.g.][]{BCGChiu}.} The reason this is not explored here is that the data in the centre of the cluster ($\lessapprox 30~{\rm kpc}$) are quite poor. A better study might be to use a galaxy cluster sample with strong-lensing data such as  \cite{CLASH} to model the BCG more accurately. The accuracy of the central data might be increased as the assumption of hydrostatic equilibrium is not required for lensing.

\begin{figure*}
\centering
\begin{tabular}{cc}
\includegraphics[scale=0.7]{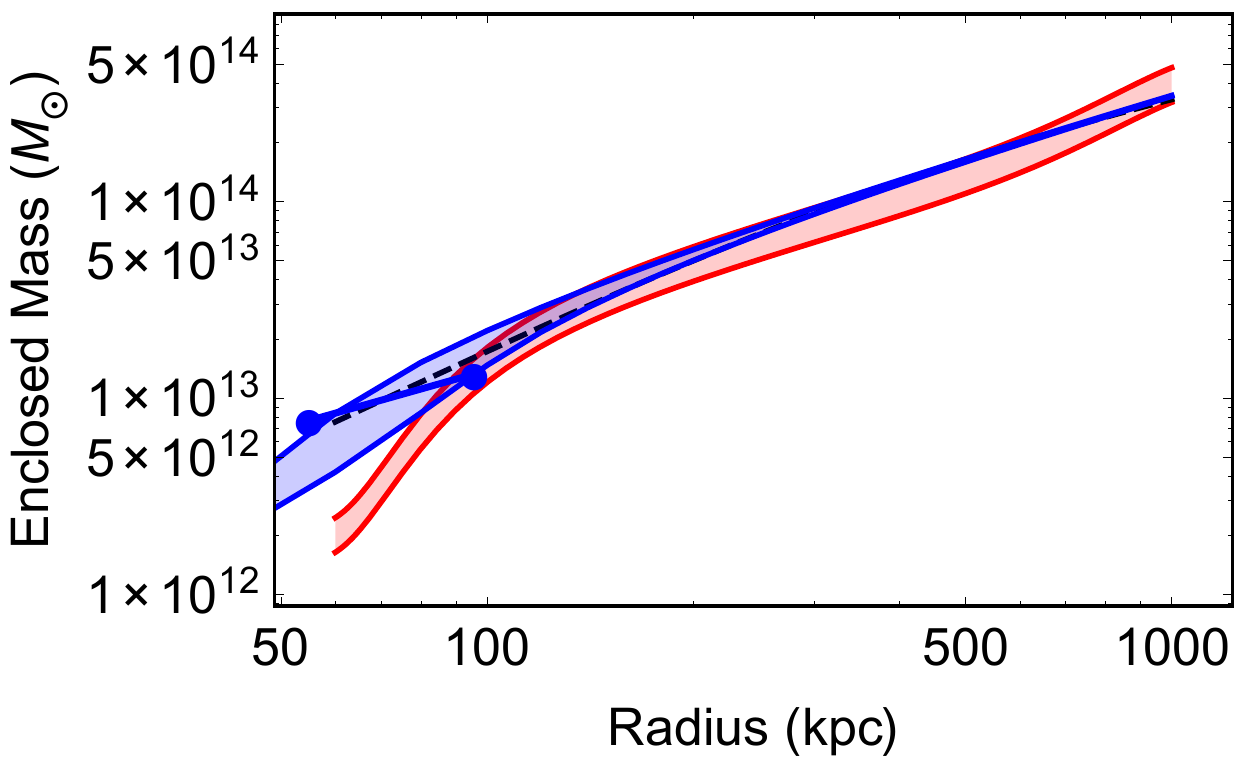} & \includegraphics[scale=0.7]{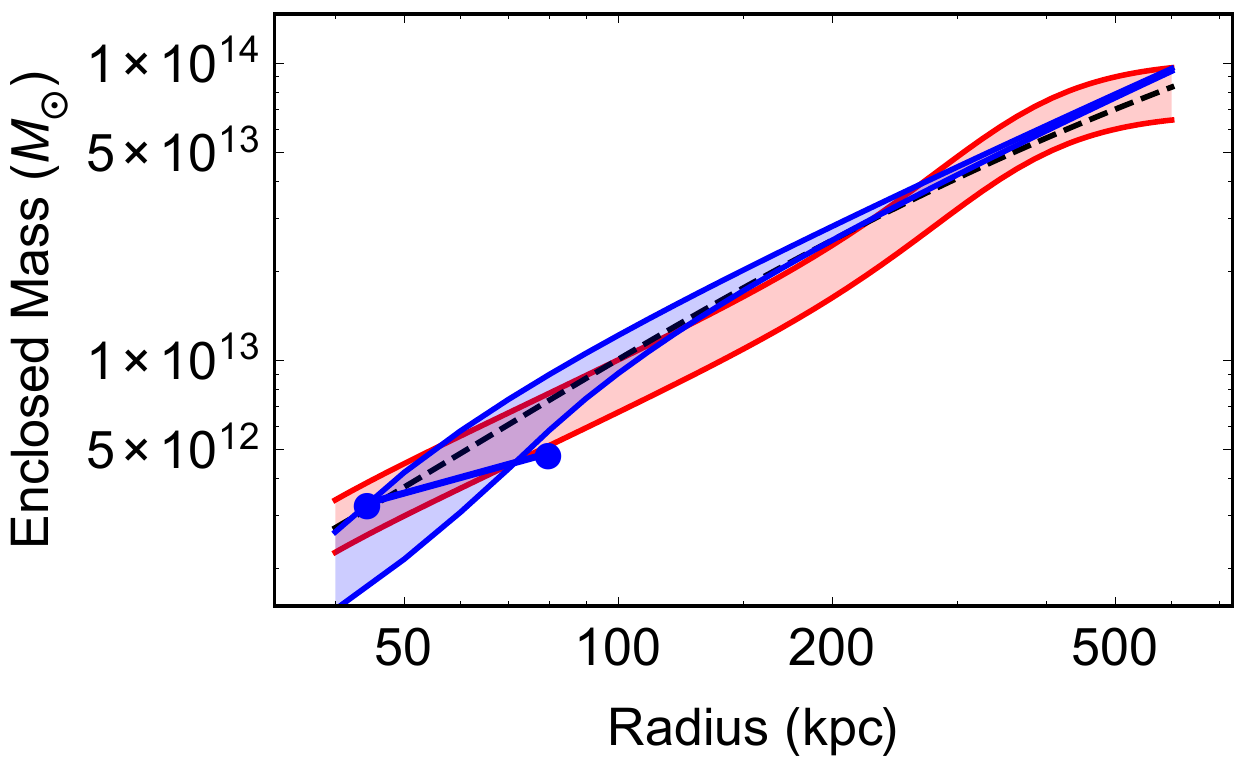} \\
\includegraphics[scale=0.7]{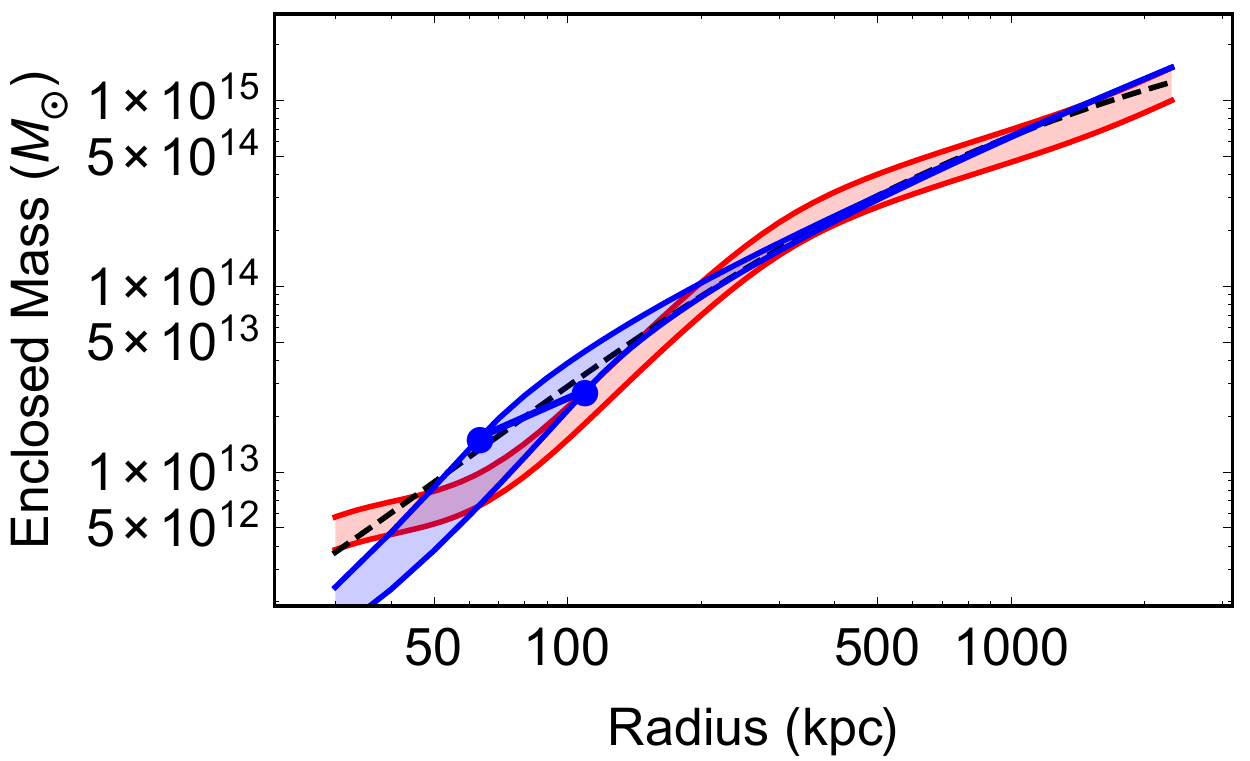} & \includegraphics[scale=0.7]{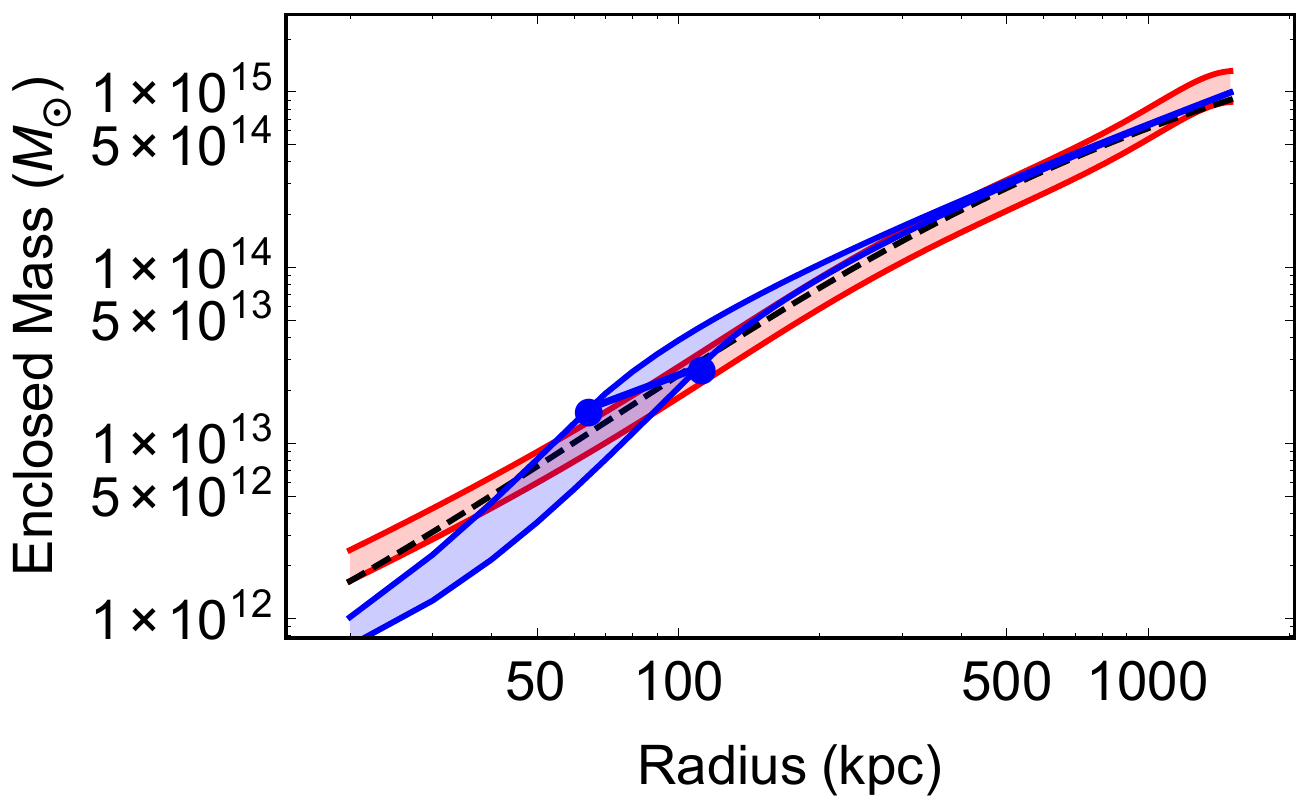}\\
\end{tabular}
\caption{Enclosed mass profiles for the cluster sample. Red shaded regions show the $\pm 20\%$ values of the dynamical mass calculated via Eq.~\ref{hydrostaticequi}. The blue shaded region shows the mass profile of the SfDM paradigm to highlight the  $\Lambda m^{3}$ dependence (upper and lower bands represent a larger and smaller choice of  $\Lambda m^{3}$ , respectively). We chose parameters  $\Lambda m^{3} = (0.1-0.3)$~$\times~ 10^{-3}$ eV$^{4}/c^{8}$. The black dashed line represents the NFW profile as given in \cite{sample}. Blue circles show the superfluid core radii for the upper and lower choices of $\Lambda m^{3}$. The best-fit analytical gas and temperature profiles did not force monotonically increasing dynamical masses. To account for this, we removed unphysical features from the plot. Top left: A133 with a virial mass of $6 \times 10^{14} M_{\odot}$, top right: A262 with a
virial mass of $2 \times 10^{14} M_{\odot}$, bottom left: A478 with a virial mass of $1.5 \times 10^{15} M_{\odot}$ , and botton right: A1413 with a virial mass of $1.5 \times 10^{15} M_{\odot}$. }
\label{A133mass}
\end{figure*}

\begin{figure*}
\centering
\begin{tabular}{cc}
\includegraphics[scale=0.7]{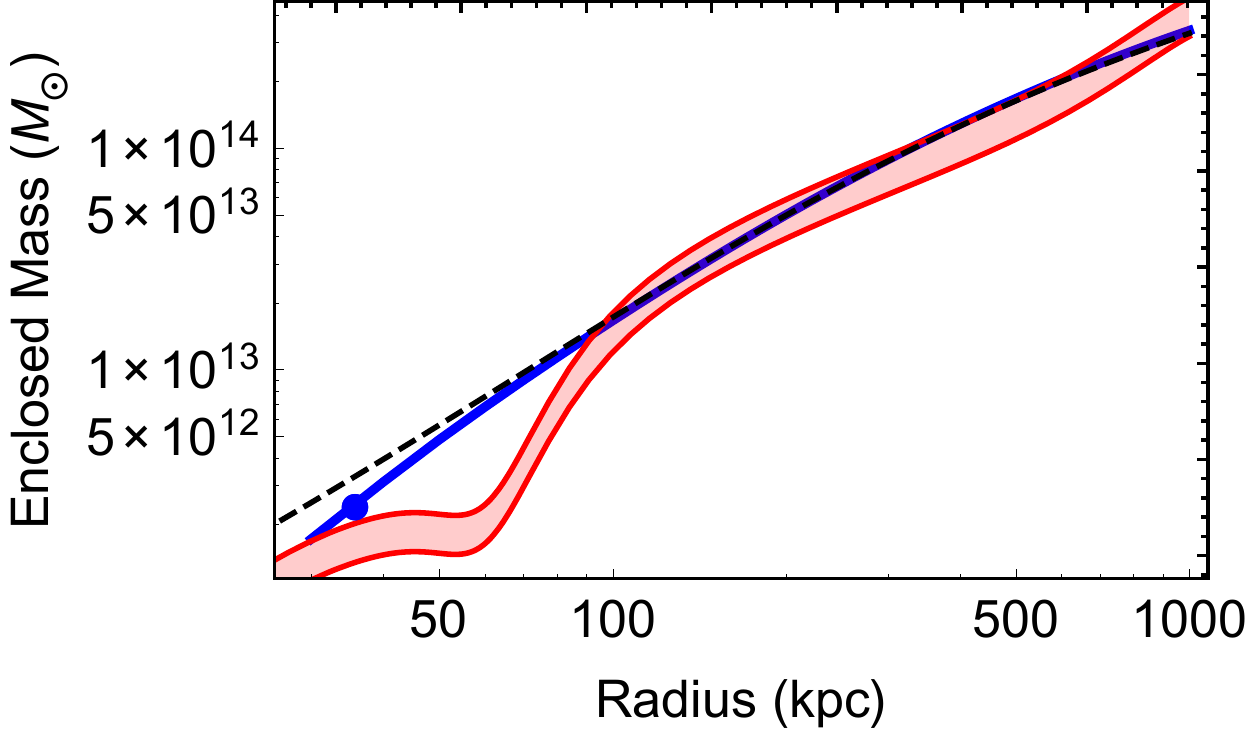} & \includegraphics[scale=0.7]{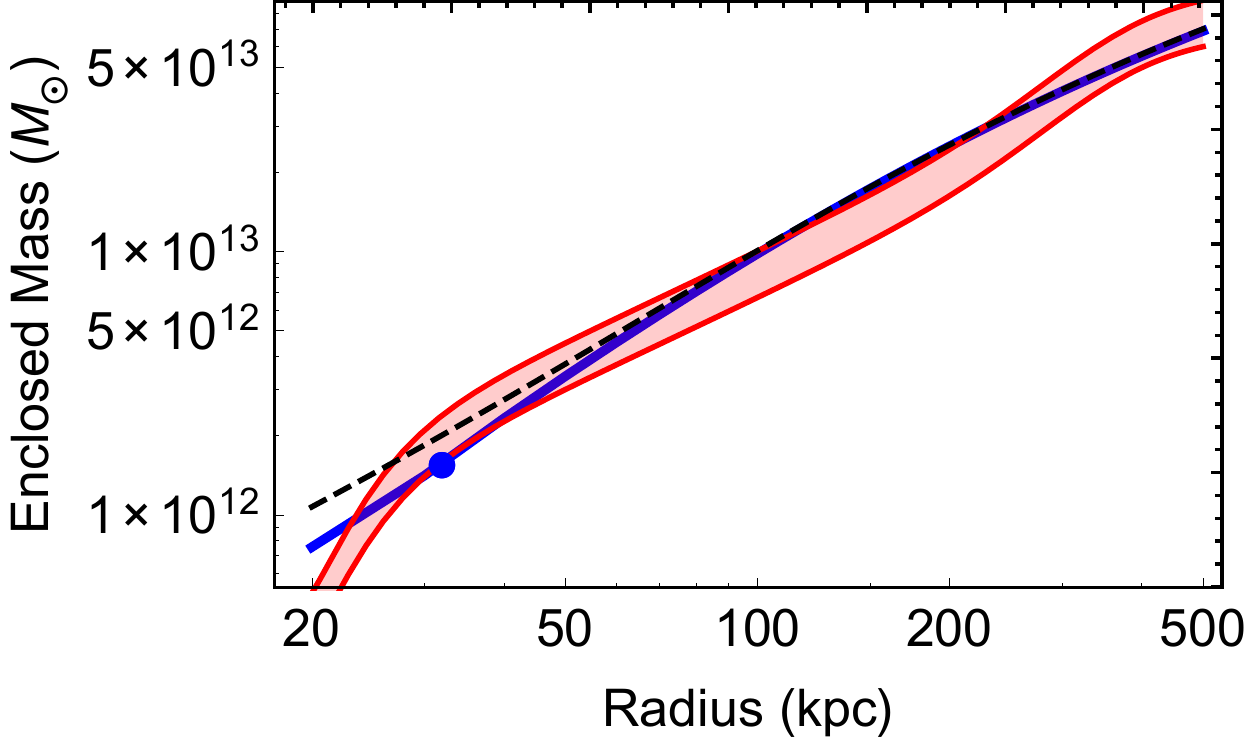} \\
\includegraphics[scale=0.7]{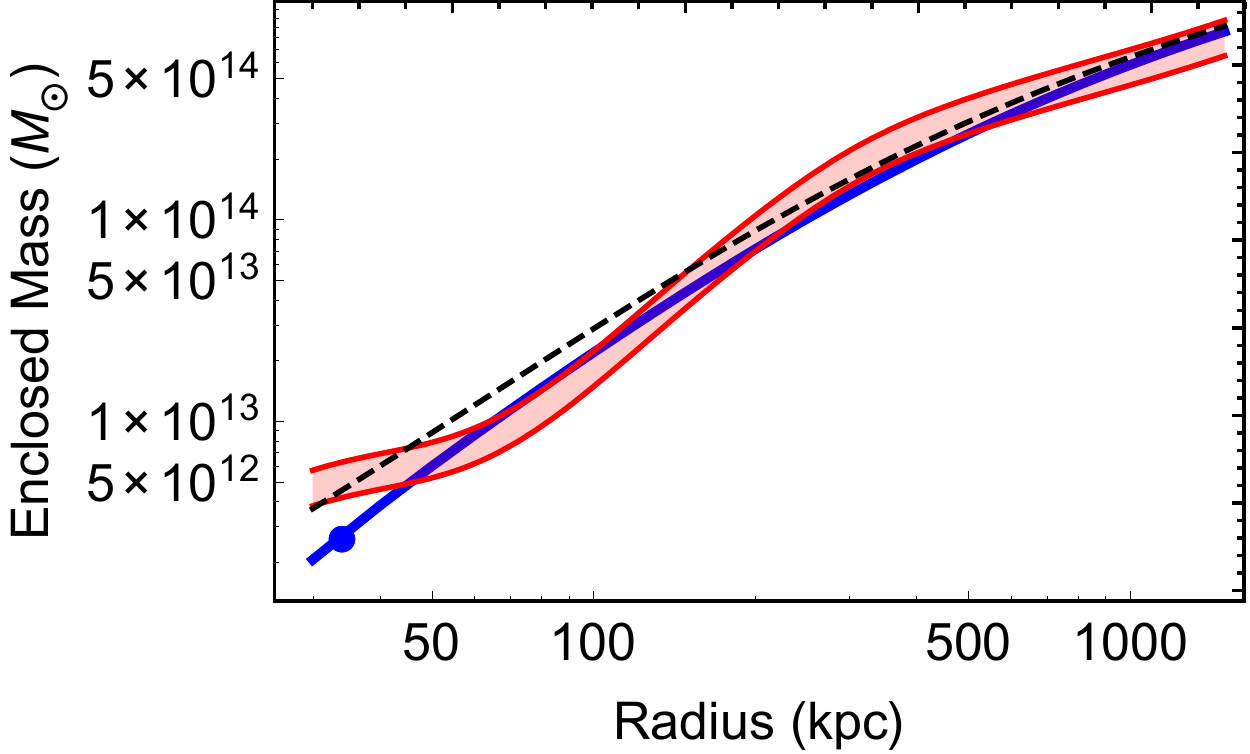} & \includegraphics[scale=0.7]{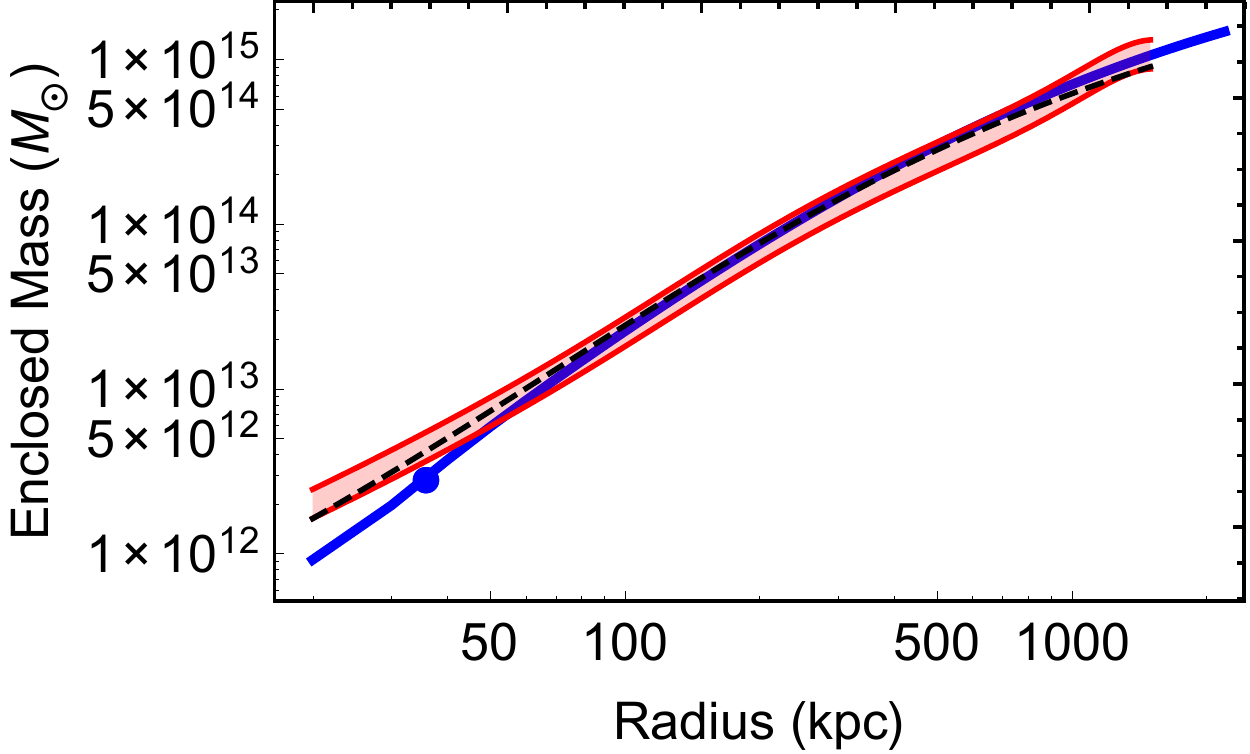}\\
\end{tabular}
\caption{Same as Fig.~\ref{A133mass} for the NFW normal phase. Only the case of  $\Lambda m^{3} = 0.3$~$\times~ 10^{-3}$ eV$^{4}/c^{8}$ is shown. $r_{s}$ for each cluster is defined as a fraction $1/n$ of the virial radius, where $n$ is 5, 6, 4, and 4 for clusters A133, A262, A468, and A1413, respectively. \textcolor{black}{The NFW normal phase follows the $\Lambda$CDM prediction much more closely than the isothermal normal-phase model. Central regions are still consistently smaller than in the $\Lambda$CDM predictions, which needs to be investigated in further work.}}
\label{NFWmass}
\end{figure*}

The derived mass profiles are plotted in Fig.~\ref{A133mass} for the isothermal normal phase and in Fig.~\ref{NFWmass} for the NFW normal phase. 

\noindent In these plots, we compare the superfluid mass (blue shaded region for Fig.~\ref{A133mass} and blue line for Fig.~\ref{NFWmass}) to the dynamical mass (red shaded region). We also show the best-fit NFW DM profile (black dashed line) as given in \citet{sample} to compare the $\Lambda$CDM result to the superfluid result. In practice, the best $\Lambda$ and $m$ parameters should be
determined for the theory, but this needs to be done in accordance with galaxy rotation curves. This is not attempted in this work.

\subsection{Isothermal analysis}

We first discuss the case where an isothermal normal phase is imposed. Figure \ref{A133mass} shows that our density matching is approximately in the range of 50-100 kpc for the clusters, depending on our choice of $\Lambda m^{3}$. Beyond this, in our normal phase, the mass profile is not perfectly in conjunction with the data, which is mainly due to the simplicity of the isothermal profile. Moreover, we have assumed that no normal phase matter is present below our pressure-matching radius. We imposed this because of modelling constraints at this stage, but future work should be conducted into how representative this assumption is compared to simulation of a cluster. \textcolor{black}{Khoury and collaborators are currently working on this problem.} Finally, although small, the phonon force might increase the mass budget slightly. Quantifying this is the next challenge.{ In short, with the limitations of the model in mind, the main mismatch between data and theory is the constant-density core. In addition, the outer parts of the clusters should be better described by an NFW profile, as in the next subsection.}

%We see from Figure ~\ref{A133mass} how the choice of $\Lambda m^{3}$ affects the mass profile. In 
%these plots, an increasing mass corresponds to an increasing value for $\Lambda m^{3}$, for which we %choose values ranging from $\Lambda m^{3} = (0.1-0.3)$~$\times~ 10^{-3}$ eV$^{4}/c^{8}$. If, for
%example, the mass of the DM particle was fixed at $0.6$ eV, this would correspond to $\Lambda \approx 
%(0.4-1)$~$\times~ 10^{-3}$ eV$/c^{2}$.What we find is that for an increasing value of $\Lambda m^{3}$,
%the core radius significantly decreases. It then follows that as the clusters tend to prefer a larger value of the
%$\Lambda m^{3}$, the clusters prefer to have a small core radius with the majority of the cluster being
%dominated by the normal phase DM. Future work with galaxies and rotation curves should provide further
%constraints on this parameter choice.

\subsection{NFW normal phase}

Owing to the simplicity of the isothermal model, other normal-phase profiles need to be tested. A natural choice is one that mimics the behaviour of the NFW profile (see Eq.~\ref{NFWDen}). We show the results for this in Fig. \ref{NFWmass}. For this figure, we only show the results for the higher value of $\Lambda m^{3}$ from our isothermal analysis. We can immediately see two consequences of imposing this normal-phase component. Firstly, the outer profile \textcolor{black}{is more in accordance with the data than the isothermal profile}, as expected. The second consequence is that as expected from Fig. \ref{Rcprofile}, the pressure matching occurs at a smaller radius. Therefore, for the same choice of $\Lambda m^{3}$ parameter, the NFW normal phase halo can match the data for a wider range of radii than for the isothermal case. We must again stress that the physical size of the ``true" superfluid core (where the superfluid density drops to zero) would be the same in both cases.

\subsection{Discrepancy in the centre}

In $\Lambda$CDM, as mentioned, the commonly used DM profile is the NFW model, which is cuspy in the centre. Because a superfluid phase is included, the DM in our model has a cored profile. Therefore, to be consistent, the cuspy-ness in our model is achieved by the stellar BCG component. The BCG mass is not a well-constrained property, hence the approximate model we use. We can conclude,
however, that if the BCG mass is much lower in the central regions than we have estimated, a challenge to the superfluid description may arise. \textcolor{black}{It is clear from Fig.~\ref{NFWmass} that the constant-density superfluid core results in a systematic feature, the superfluid mass is lower than that predicted by $\Lambda$CDM. This could be rectified by using a slightly larger BCG, but this would need to be justified observationally. We also need to determine whether this systematic feature could be permitted within typical galaxy cluster mass error bars.} 

\section{Discussion and conclusion}\label{conclusion}

We have outlined a procedure for calculating spherical cluster models, embedded in a DM halo consisting of a superfluid. The simplified model assumes that DM exists in two distinct phases: a superfluid core, and a normal-phase halo. We did not take into consideration a possible transition region with a mixture of superfluid and normal-phase DM.  This is left for further work. We tested two types of normal components: an isothermal profile, and and NFW-like profile. The NFW was shown to produce better fits than the isothermal profile, although both models show a tendency to under-predict the mass in the centre in some cases
because of the constant DM density core inherent to our present modelling. We stress that we also assumed that the phonon contribution to the force is negligible. This is a valid approximation as the phonon contribution adds a MOND-like force, which is known to be comparatively weak in clusters. This is certainly not valid in galaxies. However, the phonon contribution to the mass profile in the centre could play a role.

We applied our model to a set of galaxy clusters, comparing the mass derived from hydrostatic equilibrium and that of our superfluid model. We did not attempt to make a rigorous error analysis and $\chi^{2}$ fitting of the superfluid mass and the parameters that describe the superfluid, however. This needs to be done in conjunction with galaxies. Modelling galaxies is a much more complicated procedure because it must understood how the phonon contribution is to be modelled, and the spherical symmetry assumptions may need to be relaxed in order to correctly model disks. It is for these reasons that galaxies should be left for further analysis (Berezhiani et al. in prep.). { We can conclude from the present analysis, however, that when a much lower value of $\Lambda m^{3}$ is  needed in galaxies, the constant-density core is extended in clusters, and this increase the discrepancy with data.}

There is indeed some discrepancy between the superfluid result and the dynamical mass estimates of the cluster. For lower values of $\Lambda m^{3}$, { the superfluid core seems to systematically under-predict the mass profile in the galaxy cluster centre}. This is a result of the superfluid recipe predicting a constant-density core, whereas the data seem to prefer a cuspier central profile. This needs to be rigorously tested against strong-lensing data of galaxy clusters as an independent test. We did achieve better fits with a larger choice of $\Lambda m^{3}$ at the cost of a smaller core radius, which could be in contention with galaxy data. However, as we mentioned, the true extent of the superfluid core will be larger than our quoted values, so this may not be an issue, and adding a phonon contribution to the centre might slightly alleviate the discrepancy.

The next goal is to perform lensing tests with galaxy clusters in the superfluid paradigm as an independent test. When this is completed, we may move on to model galaxies, with an understanding of how the phonon force can be included. In order to do this, we need to construct a more complicated model that includes a transition region between the superfluid and normal phase. { With all these caveats in mind, the present work nevertheless gives an order-of-magnitude estimate of the parameter values of the theory that do not violently disagree with cluster data. If much lower values of $\Lambda m^{3}$ were  needed in galaxies, it would be difficult to reconcile the superfluid DM framework with galaxy clusters.}

\section*{Acknowledgements}

We thank Lasha Berezhiani for helpful discussions. We also thank Alexey Vikhlinin for helpful correspondence on the cluster data. We warmly thank Priya Natarajan and Doug Finkbeiner for hosting a stimulating Radcliffe Exploratory Seminar at Harvard University where this collaboration was initiated. We would also like to thank the anonymous referee for their helpful comments on style and content that improved the paper. AOH is supported by Science and Technologies Funding Council (STFC) studentship (Grant code: 1-APAA-STFC12). J.K. is supported in part by NSF CAREER Award PHY-1145525, NASA ATP grant NNX11AI95G, and the Charles E. Kaufman Foundation of the Pittsburgh Foundation. BF acknowledges financial support from the ``Programme Investissements d'Avenir'' (PIA) of the IdEx from the Universit\'e de Strasbourg. 
\bibliographystyle{aa} % style aa.bst
\bibliography{superfluidbib} % your references Yourfile.bib

%\begin{figure*}
%\centering
%\begin{tabular}{cc}
%\includegraphics[scale=0.6]%{A133_Mass_parameter_choice1.pdf} & %\includegraphics[scale=0.6]%{A133_Mass_parameter_choice2.pdf}\\
%\includegraphics[scale=0.6]%{A262_Mass_parameter_choice1.pdf} & %\includegraphics[scale=0.6]%{A262_Mass_parameter_choice2.pdf}\\
%\includegraphics[scale=0.6]%{A478_Mass_parameter_choice1.pdf} & %\includegraphics[scale=0.6]%{A478_Mass_parameter_choice2.pdf}\\
%\includegraphics[scale=0.6]%{A1413_Mass_parameter_choice1.pdf} & %\includegraphics[scale=0.6]%{A1413_Mass_parameter_choice2.pdf}
%\end{tabular}
%\caption{}
%\label{}
%\end{figure*}

\end{document}